\documentclass[]{iopart}
\pdfoutput=1
\usepackage{graphicx}
\usepackage{amssymb}
\usepackage{iopams}
\usepackage{setstack}
\usepackage[utf8]{inputenc}
\usepackage{cancel}
\usepackage[usenames,dvipsnames]{color}
\usepackage{psfrag}
\usepackage{epsfig}
\usepackage{bbm}
\usepackage{bm}
\usepackage{dsfont}
\usepackage{soul}
\usepackage{colonequals}
\usepackage{float}
\usepackage{enumitem}
\usepackage{cite}
\usepackage{setspace}
\usepackage{cases}
\usepackage[normalem]{ulem}

\begin{document}
	
	\title{Clustering and flocking of repulsive chiral active particles with non-reciprocal couplings} 
	\author{Kim L. Kreienkamp$^1$ and Sabine H.~L.~Klapp$^1$}
	\address{$^1$ Institut f\"ur Theoretische Physik,
		Hardenbergstr.~36,
		Technische Universit\"at Berlin,
		D-10623 Berlin,
		Germany}

	\begin{abstract}
		Recently, non-reciprocal systems have become a focus of growing interest. Examples occur in soft and active matter, but also in engineered quantum materials and neural (brain) networks. Here, we investigate the impact of non-reciprocity on the collective behavior of a system of (dry) chiral active matter. Specifically, we consider a mixture of ``circle swimmers'' with steric interactions and non-reciprocal alignment couplings. 
		Based on hydrodynamic equations which we derive from a set of Langevin equations, we explore the interplay of non-reciprocity, finite size, and chirality.
		We first consider, as a reference, one-species systems with reciprocal couplings. Based on a linear stability analysis and numerical simulations, we here observe three different types of collective behavior, that is, flocking, motility-induced phase separation, and a combination of both. 
		Turning then to a non-reciprocal system, we find that non-reciprocity can turn otherwise stationary instabilities into oscillatory ones, affect the relative orientation of flocks, and, crucially, change the general type of instability.  
		This illustrates the drastic impact of non-reciprocity on the emergent collective dynamics of chiral active matter systems, with potentially far-reaching biological implications.
	\end{abstract}
	\noindent{\it Keywords\/}: non-reciprocity, chiral active particles, motility-induced phase separation, flocking, hydrodynamic field equations
	
	\maketitle
	\onehalfspacing
	
	\section{Introduction}
	\label{sec:introduction}
	Recently, the physics of interacting systems with non-reciprocal couplings has become a focus of growing interest. In soft-matter systems, non-reciprocity occurs, e.g., when inter-particle forces are mediated by a non-equilibrium environment \cite{Bowick_2022_symmetry_thermodynamcs_topology_active_matter,You_Baskaran_Marchetti_2020_pnas,scheibner_2020_odd_elasticity,gupta_2022_non-reciprocity_elastic_medium}, yielding hydrodynamic or phoretic interactions \cite{saha_scalar_active_mixtures_2020}. Further, non-reciprocity can occur in heterogeneous, multi-component systems like bacterial suspensions \cite{xiong_2020_flower-like_patterns_bacteria,theveneau_2013_chase-and-run_cells}, mixtures of diffusiophoretic colloids \cite{Saha_2019_chemotactive_active_colloids,soto_2014_self-assembly_catalytically_active_colloidal_molecules}, neural (brain) networks \cite{sompolinsky_1986_asymmetric_neural_networks,brunel_2000_dynamics_neural_network,golomb_2000_neural_network_random_coupling,borgers_2003_synchronization_networks}, prey-predator systems \cite{tsyganov_2003_predator-prey_system,meredith_2020_predator--prey_droplets}, and social groups \cite{helbing_1995_social_forces_pedestrians,hong_2011_kuramoto_model_positive_negative_coupling_parameters}. Additional examples of non-reciprocal systems include ``cognitive'' particles or agents with a vision cone \cite{Barberis_2016_cognitive_flocking_model,loos_non-reciprocal_xy_model_vision_cone,Lavergne_2019_group_formation_visual_perception-dependent_motility}, as well as engineered quantum materials \cite{metelmann_2015_nonreciprocal_photon_transmission,zhang_2018_non-reciprocal_quantum_optical_system,mcdonald_2022_nonequilibrium_quantum_non-Hermitian_lattice}. In contrast to equilibrium systems governed by Newton's third law, non-reciprocal systems are generally considered to be out of equilibrium \cite{Ivlev_2015_statistical_mechanics_where_newtons_third_law_is_broken,Loos_2020_Irreversibility_non-reciprocal_interactions}.
	
	Given their non-equilibrium character and ubiquity in nature, various recent studies \cite{Bowick_2022_symmetry_thermodynamcs_topology_active_matter,fruchart_2021_non-reciprocal_phase_transitions,You_Baskaran_Marchetti_2020_pnas,saha_scalar_active_mixtures_2020,frohoff-hulsmann_2021_localized_states,frohoff-hulsmann_2021_Cahn-Hilliard_oscillatory} have addressed the collective dynamics of non-reciprocal (soft matter) systems based on a field-theoretical approach. While the detailed effect of non-reciprocity depends on the system considered, an overall finding is that non-reciprocity can drive time-dependent states.
	In fact, already for a two-component system of purely diffusive, conserved scalar fields, You et al.~\cite{You_Baskaran_Marchetti_2020_pnas} have shown that non-reciprocity constitutes a generic route to traveling states. They found that a static demixed pattern can undergo a transition to a spatially inhomogeneous ``run-and-catch'' state, which breaks parity and time-reversal symmetry. 
	
	In the present work, we focus on \textit{active} systems, whose constituents perform persistent motion due to an internal or external source of energy. Thus, even in the conventional case of reciprocal couplings, these systems are intrinsically out of equilibrium. It is now well established that active systems are capable of exhibiting a variety of non-equilibrium phase transitions and self-organization without external driving \cite{marchetti_2013_hydrodynamics_soft_active_matter,ramaswamy_2017_active_matter}.
	A growing number of studies is exploring the impact of \textit{non-reciprocal alignment} between the active constituents. 
	Examples include systems of (passive) dissenters in a flock of active particles \cite{yllanes_2017_dissenters_to_disorder_a_flock} and generic phase transitions in non-reciprocal, active, two-species systems \cite{fruchart_2021_non-reciprocal_phase_transitions,frohoff-hulsmann_2021_localized_states,frohoff-hulsmann_2021_Cahn-Hilliard_oscillatory}.
	In particular, \cite{fruchart_2021_non-reciprocal_phase_transitions} demonstrated that non-reciprocity alone can destabilize the stationary (anti-)flocking state, characterized by (anti-)parallel motion of the particles of both species. The related phase transition is marked by an exceptional point in the space of field variables, resulting in a time-dependent, so-called ``chiral phase''. Here, flocks of both species rotate at a constant speed with a fixed relative angle, although there is no intrinsic torque on the particle level.
	
	Given these recent developments, the goal of our work is to examine the \textit{combined} effect of two ubiquitous features of active matter systems, namely non-reciprocity and chirality of individual particles.
	Specifically, we consider a \textit{mixture} of chiral active particles with non-reciprocal \mbox{(anti-)}alignment between particles of different species and mutual repulsion. In contrast to conventional ``linear'' swimmers that change their direction of motion only by diffusion or orientational (alignment) interactions (such as, e.g., active Brownian particles \cite{romanczuk_2012_active_brownian_particles,buttinoni_2013_dynamical_clustering} or Vicsek particles \cite{vicsek_1995_flocking,vicsek_collective_motion_2012}), chiral active particles (sometimes also named as ``circle swimmers'') additionally self-rotate with an intrinsic frequency \cite{loewen_2016_chirality_microswimmer_motion,van_teeffelen_2008_single_circle_swimmer}. The intrinsic rotation can be caused, e.g., by a chiral body shape as in anisotropic colloids \cite{campbell_2017_helical_paths_anisotropic_colloid}, curved proteins \cite{loose_2014_curved_proteins} or artificial L-shaped particles \cite{kummel_2013_L-shaped_particle,ten_Hagen_2014_gravitaxis_L-shaped_particle}. In two dimensions, a chiral body shape indeed leads to a circular motion \cite{loewen_2016_chirality_microswimmer_motion,ledesma_2012_circle_swimmer_two_dimensions_low_Re,van_teeffelen_2008_single_circle_swimmer}. Other examples of chiral active particles are E.~coli bacteria close to walls and interfaces \cite{di_leonardo_2011_swimming_bacteria_interface,lauga_2006_circle_swimming_bacteria,berg_1990_bacteria_swimming_circle,maeda_1976_circle_swimming_bacteria}, sperm cells \cite{friedrich_2007_chemotaxis_sperm_cells,riedel_2005_vortex_sperm_cell}, and particles actuated by rotating fields \cite{mano_2017_Janus_particle_electric_field,erglis_2007_magnetotactic_bacteria_rotating_field,cebers_2011_diffusion_magnetotactic_bacteria,tierno_2021_transport_assembly_magnetic_rotors}.
	
	To examine the combined effect of non-reciprocity and chirality on the collective dynamics, we derive hydrodynamic equations for the density and polarization fields starting from the microscopic Langevin equations governing the motion of individual particles. We thereby employ a mean-field approximation and a truncation scheme to get rid of higher-order moments. While this strategy has been used before \cite{fruchart_2021_non-reciprocal_phase_transitions,Farrell_2012_Pattern_formation_self-propelled_particles_density-dependent_motility,yllanes_2017_dissenters_to_disorder_a_flock}, including applications for circle swimmers \cite{Liebchen_2016_pattern_formation_chemically_interacting_active_rotors,liebchen_2017_collective_behavior_chiral_active_matter_pattern_formation_flocking, Levis_2019_activity_induced_synchronization}, our approach additionally takes into account the impact of repulsive interactions.
	
	It is well established that steric repulsion leads to motility-induced phase separation already in ``simple'' active fluids \cite{Cates_2013_density-dependent_velocity_MIPS,Bialke_2013_microscopic_theory_phase_seperation,speck_2015_dynamical_mean_field_phase_separation,damme_2019_interparticle_torques_phase_separation,worlitzer_2021_mips_meso-scale_turbulence_active_fluids,elena_2018_velocity_alignment_MIPS,elena_2021_phase_separation_self-propelled_disks,Liao_2018_circle_swimmers_monolayer,liao_2020_dynamical_self-assembly_dipolar_active_Brownian_particles,barre_2015_MIPS_velocity_alignment,gonnella_MIPS_2015}. A further well-known fact is that systems with strong alignment couplings exhibit a flocking transition towards large-scale ordered motion (e.g., \cite{vicsek_1995_flocking,Czirok_1997_ordered_motion,toner_tu_1995_long_range_order,toner_tu_2005_hydrodynamics_phases_of_flocks,liebchen_2017_collective_behavior_chiral_active_matter_pattern_formation_flocking,Gregoire_2004_collective_motion,Chate_2008_collective_motion,marchetti_2013_hydrodynamics_soft_active_matter,Solon_2013_flocking_transition}). With this in mind, we here address the question of how these phenomena are affected by the combination of intrinsic rotation frequency and non-reciprocal orientational couplings. To this end, we combine an analytical linear stability analysis and numerical continuum simulations of the full, non-linear hydrodynamic equations, which allow for relatively quick explorations of interesting parameter regimes.
	
	The paper is organized as follows. In \sref{sec:model}, we start by introducing the microscopic model, followed by the derivation of the hydrodynamic equations. In \sref{sec:one_species}, we first consider a one-species chiral system with reciprocal alignment couplings, where we examine different scenarios of collective behavior and the impact of intrinsic rotation. We then turn to a two-species system with non-reciprocal couplings in \sref{sec:two_species}. Considering selected values of the intrinsic frequencies, we study in \sref{sec:two_species}, how non-reciprocity affects the collective behavior of the chiral mixture. We close with a discussion of our results in \sref{sec:conclusion}.

	\section{Model}
	\label{sec:model}
	
	\subsection{Particle-level description}
	We consider a two-dimensional system of chiral active particles comprising two species $a=A,B$. The $N = \sum_a N_a$ particles are located at positions $\bm{r}_{\alpha}$ (with $\alpha=i_a = 1,...,N_a$) and move like active Brownian particles (ABP) with additional intrinsic torque. Thus, they rotate with a species-specific intrinsic frequency $\omega_a$, and self-propel with velocity $v_a$ along the instantaneous direction $\bm{p}_{\alpha}(t)=({\rm{cos}}\,\theta_{\alpha}, {\rm{sin}}\,\theta_{\alpha})^{\rm{T}}$, where $\theta_{\alpha}$ is the polar angle. The dynamics is then given by the overdamped Langevin equations (LE)
	\numparts
	\begin{eqnarray}
	\label{eq:Langevin_eq}
		\dot{\bm{r}}_{\alpha}(t) = v_a\,\bm{p}_{\alpha}(t) + \mu_r \, \sum_{\beta\neq\alpha} \bm{F}_{\rm{sr}}(\bm{r}_{\alpha},\bm{r}_{\beta}) + \bm{\xi}_{\alpha}(t) \label{eq:Langevin_r}\\
		\dot{\theta}_{\alpha}(t) = \omega_{{a}} + \mu_{\theta} \, \sum_{\beta\neq\alpha} \mathcal{T}_{\rm al}^{ab}(\bm{r}_{\alpha},\bm{r}_{\beta},\theta_{\alpha},\theta_{\beta}) + \eta_{\alpha}(t) \label{eq:Langevin_theta},
	\end{eqnarray}
	\endnumparts
	where the sums over particles $\beta=j_b=1,...,N_b$ couple the dynamics of particle $\alpha$ to the position and orientation of all other particles of both species $b=A,B$.
	
	The translational LE \eref{eq:Langevin_r} involves the soft repulsive force 
	\begin{equation}
	\label{eq:steric_repulsion}
		\bm{F}_{\rm{sr}}(\bm{r}_{\alpha},\bm{r}_{\beta}) = \frac{\bm{r}_{\alpha\beta}}{r_{\alpha\beta}} \, k \, (R_{r} - r_{\alpha\beta}) \, \Theta(R_{\rm r} - r_{\alpha\beta})
	\end{equation}
	within a cutoff-distance $R_{r}$, where $r_{\alpha\beta} = \vert \bm{r}_{\alpha\beta}\vert = \vert \bm{r}_{\alpha}-\bm{r}_{\beta} \vert$ and $\Theta(R_{r} - r_{\alpha\beta}) = 1$, if $r_{\alpha\beta}<R_r$, and zero otherwise. Note that, for simplicity, we assume that steric repulsion of strength $k$ is the same for all particles (i.e.~$k$ and $R_{r}$ are species-independent). Further, we here assume a soft, piece-wise linear repulsive force \cite{yllanes_2017_dissenters_to_disorder_a_flock,You_Baskaran_Marchetti_2020_pnas}, which allows for an analytical treatment in our continuum description (see \sref{ssec:coarse-grained_description}). 
	
	The rotational LE \eref{eq:Langevin_theta} contains the torque, whose in-plane component is given by
	\begin{equation}
	\label{eq:torque}
		\mathcal{T}_{\rm al}^{ab}(\bm{r}_{\alpha}, \bm{r}_{\beta}, \theta_{\alpha}, \theta_{\beta}) = K_{ab}\, {\rm{sin}}(\theta_{\beta}-\theta_{\alpha}) \, \Theta(R_{\theta}-r_{\alpha\beta}) ,
	\end{equation}
	of strength $K_{ab}$, which can be positive or negative. The sinusoidal dependence of the torque on the relative angle $\theta_{\beta}-\theta_{\alpha}$ is motivated by the expression derived from an orientation-dependent potential of the form $\bm{p}_{\alpha} \cdot \bm{p}_{\beta}$ \cite{peruani_2008_mean-field_velocity_alignment,elena_2018_velocity_alignment_MIPS,Farrell_2012_Pattern_formation_self-propelled_particles_density-dependent_motility,barre_2015_MIPS_velocity_alignment}. From \Eref{eq:torque} it follows that particles of species $a$ tend to orient parallel (align) or anti-parallel (anti-align) with neighboring particles (within radius $R_{\theta}$) of species $b$ when $K_{ab}>0$ or $K_{ab}<0$, respectively. For \textit{reciprocal} couplings defined by the choice $K_{ab} = K_{ba}$, particles of species $a$ align (or anti-align) with particles of species $b$ in the same way as particles of species $b$ with particles of species $a$. This is, in fact, the natural choice when the orientational coupling in \Eref{eq:torque} is derived from a Hamiltonian, i.e.~a many-body interaction potential. In the present work, we specifically allow for \textit{non-reciprocal} orientational couplings, that is, $K_{ab}\neq K_{ba}$.
	
	Both the position and orientation of the particles are subject to thermal noise, modeled as Gaussian white noise processes $\bm{\xi}_{\alpha}$(t) and $\eta_{\alpha}(t)$ of zero mean and variances $\langle \xi_{\alpha,k}(t) \xi_{\beta,l}(t') \rangle = 2\,\xi\,\delta_{\alpha\beta}\,\delta_{kl} \,\delta(t-t')$ and $\langle \eta_{\alpha}(t) \eta_{\beta}(t') \rangle = 2\,\eta\,\delta_{\alpha\beta}\,\delta(t-t')$, respectively. The mobilities are connected to thermal noise via $\mu_{r} = \beta\,\xi$ and $\mu_{\theta} = \beta \, \eta$, where $\beta^{-1}=k_{\rm B}\,T$ is the thermal energy with Boltzmann's constant $k_{\rm B}$ and temperature $T$. 
	
	Finally, we specify the propulsion velocity $v_a$ of individual particles appearing in LE \eref{eq:Langevin_r}. In standard ABP-like models, one usually assumes that every individual particle self-propels with a constant speed $v_0$. It turns out, however, that a constant self-propulsion speed combined with the mean-field approximation in our continuum approach does not reproduce the motility-induced phase separation (MIPS) into low- and high-density regions characteristic for ABPs \cite{buttinoni_2013_dynamical_clustering,van_der_linden_2019_interrupted_mips,bauerle_2018_self-organization_quorum_sensing,liu_2019_self-driven_phase_transitions,OByrne_2021_introduction_MIPS}. The underlying reason is that the mean-field approach disregards the structure of pair correlations, which become anisotropic due to activity. More specifically, previous numerical \cite{Bialke_2013_microscopic_theory_phase_seperation,elena_2021_phase_separation_self-propelled_disks} and analytical \cite{Bialke_2013_microscopic_theory_phase_seperation,speck_2015_dynamical_mean_field_phase_separation, damme_2019_interparticle_torques_phase_separation,elena_2021_phase_separation_self-propelled_disks} studies have shown that the force imbalance introduced by the self-propulsion of particles (there are more particles in front of the reference particle than behind it) causes an effective velocity reduction depending on the density of surrounding particles. In order to account for this effect, we replace the (species-dependent) constant speed $v_0^a$ with an \textit{effective} density-dependent velocity of an active particle in an interacting repulsive system \cite{Bialke_2013_microscopic_theory_phase_seperation,speck_2015_dynamical_mean_field_phase_separation,damme_2019_interparticle_torques_phase_separation,Cates_2013_density-dependent_velocity_MIPS,Farrell_2012_Pattern_formation_self-propelled_particles_density-dependent_motility,elena_2021_phase_separation_self-propelled_disks,worlitzer_2021_mips_meso-scale_turbulence_active_fluids}. Specifically, we assume that the particles self-propel with effective velocity \cite{Bialke_2013_microscopic_theory_phase_seperation,speck_2015_dynamical_mean_field_phase_separation,damme_2019_interparticle_torques_phase_separation,elena_2021_phase_separation_self-propelled_disks,worlitzer_2021_mips_meso-scale_turbulence_active_fluids}
	\begin{equation}
		v_a = v^{\rm eff}_a(\rho)=v^a_0 - \zeta \, \rho .
	\end{equation}
	This choice expresses the fact that particles are slowed down in crowded situations, depending on the (species-independent) overall local particle density $\rho=\rho(\bm{r})=\sum_a \rho^a$ and velocity-reduction parameter $\zeta$. Note that since we assume that particles of each species experience the same steric repulsion (see \Eref{eq:steric_repulsion}), each particle is slowed down with the same ``frictional'' parameter $\zeta$, coupled to the overall density field $\rho(\bm{r})$.

	\subsection{Coarse-grained description}\label{ssec:coarse-grained_description}
	Starting from the particle-level (``microscopic'') Langevin equations \eref{eq:Langevin_r} and \eref{eq:Langevin_theta}, we derive a continuum model to study large-scale patterns and relevant mechanisms effecting the collecting behavior.
	To this end, we employ a coarse-graining strategy introduced in \cite{Dean_1996-Langevin_equation_interacting_particles}, yielding a mean-field Fokker-Planck equation for the one-particle probability density function (PDF) 
	\begin{equation}
		f^a(\bm{r},\theta,t) = \frac{1}{N_a} \sum_{i_a}^{N_a} \langle \delta(\bm{r}-\bm{r}_{i_a}(t)) \, \delta(\theta - \theta_{i_a}(t)) \rangle.
	\end{equation}
	Relating the Fourier coefficients of $f^a(\bm{r},\theta,t)$ to orientational moments, we then apply a closure relation neglecting higher-order moments to derive a hydrodynamic field description in terms of the particle density
	\begin{equation}
		\rho^a(\bm{r},t) = N_a \int_{-\pi}^{\pi} f^a(\bm{r},\theta,t)\,{\rm d}\theta
	\end{equation}
	and polarization density
	\begin{equation}
		\bm{w}^a(\bm{r},t) = N_a \int_{-\pi}^{\pi} f^a(\bm{r},\theta,t)\,\bm{p}(\theta)\,{\rm d}\theta,
	\end{equation}
	the latter measuring the overall orientation of particles at a certain position via $\bm{w}^a/\rho^a$ \cite{Bertin_2009_hydrodynamic_equations_self-propelled_particles}. Our derivation closely follows the approaches of deriving a hydrodynamic version of the Vicsek model presented in \cite{fruchart_2021_non-reciprocal_phase_transitions} and of non-repulsive, reciprocally coupled chiral active particles in \cite{Liebchen_2016_pattern_formation_chemically_interacting_active_rotors,liebchen_2017_collective_behavior_chiral_active_matter_pattern_formation_flocking, Levis_2019_activity_induced_synchronization}. Details regarding the derivation of the hydrodynamic equations are given in \ref{sec:derivation_hydrodynamic_equations_appendix}. Specifically, we find the continuity equation
	\begin{equation}
	\label{eq:continuum_eq_density}
		\partial_t \rho^a +  \nabla \cdot \bm{j}_{a} = 0
	\end{equation}
	with flux
	\begin{equation}
	\label{eq:continuity_flux}
		\bm{j}_a = v^{\rm eff}_a(\rho)\, \bm{w}^a - \sum_b \mathcal{R}_b\, \rho^a \nabla \rho^b -  D_{\rm t}\,\nabla\,\rho^a .
	\end{equation}
	The flux involves the polarization density, which evolves according to
	\begin{equation}
		\label{eq:continuum_eq_polarization}
		\eqalign{
			\fl \partial_t \bm{w}^a 
			= - \frac{1}{2} \, \nabla \,\big(v^{\rm eff}_a(\rho)\, \rho^a\big)  -\Omega_a \, \bm{w}^{a*} - \bm{w}^{a} + \sum_b g_{ab} \, \rho^a\, \bm{w}^b \\
			+  D_{\rm t}\,\nabla^2\,\bm{w}^a + \frac{v^{\rm eff}_a(\rho)}{4\,b_a} \, \nabla^2\,\Big(v^{\rm eff}_a(\rho)\,\Big\{2\,\bm{w}^a - \Omega_a\,\bm{w}^{a*} \Big\}\Big) \\
			- \sum_{b,c} \frac{2\, g_{ab}\,g_{ac}}{b_a} \, \Big[2 \, \bm{w}^a \, (\bm{w}^b \cdot \bm{w}^c) - \Omega_a \, \bm{w}^{a*} \, (\bm{w}^b \cdot \bm{w}^c) \Big] \\
			+ \sum_b \mathcal{R}_b \, \bm{w}^a\, \nabla^2 \rho^b + O(\nabla \bm{w}^2) + O(\nabla \rho \nabla \bm{w}).
		}
	\end{equation}
	In \Eref{eq:continuum_eq_polarization}, we have neglected the explicit terms of order $O(\nabla \bm{w}^2)$ and $O(\nabla \rho \nabla \bm{w})$. The full equation is shown in \ref{sec:full_hydrodynamic_equations}.
	
	The density flux $\bm{j}_a$ given in \Eref{eq:continuity_flux} reflects that the motion of particles of species $a$ in space arises from their self-propulsion in direction $\bm{w}^a$, whereby the particles are slowed down in crowded situations due to the density-dependent velocity. Additionally, the flux comprises the drift of particles towards less crowded regions due to steric repulsion and translational diffusion. 
	The change of the polarization density $\bm{w}^a$, described by \Eref{eq:continuum_eq_polarization}, originates from the competition between the tendency of particles to swim (with increasing speed) towards low-density regions (first term on r.h.s.), the rotation of the polarization with intrinsic frequency (second term), the decay of the polarization due to rotational diffusion (third term), and the orientational coupling of particles among all species (fourth term). The remaining (diffusional and non-linear) terms smear out low- and high-polarization regions.

	Further, in Equations \eref{eq:continuity_flux} and \eref{eq:continuum_eq_polarization}, we have introduced $v^{\rm eff}_a(\rho)={\rm Pe}_a- z\,\rho$ with $\rho=\sum_b\rho^b$, $b_a = 2(4+\Omega_a^2)$, $\bm{w}^*=(w_y, -w_x)^{\rm T}$, and $\nabla^*=(\partial_y, -\partial_x)^{\rm T}$. The equations are non-dimensionalized by choosing the Brownian time scale $\tau=1/\eta$ as characteristic time scale and the particle radius $\ell$ as characteristic length scale. The particle and polarization densities of species $a$ are scaled with the average particle density $\rho_0^a$. The remaining six dimensionless control parameters are the P\'eclet number ${\rm Pe}_a=v_0^a\,\tau/\ell$, $\mathcal{R}_a = 2\,k\,\mu_r\,R_{r}^4\,\pi\,\rho_0^a\,\tau/(3\,\ell^2)$ encoding the strength of the repulsive force, $z = \zeta\,\rho_0\,\tau/\ell$ measuring the particle velocity-reduction due to the environment, the translational diffusion coefficient $D_{\rm t}=\xi\,\tau/\ell^2$, $\Omega_a =\omega_a\,\tau$ for the intrinsic frequency, and $g_{ab} = K_{ab}\,\mu_{\theta}\,R_{\theta}^2\,\pi\,\rho_0^b\,\tau/2$ as relative orientational coupling parameter. Thereby, $g_{ab}>0$ leads to an alignment and $g_{ab}<0$ to an anti-alignment of particles. The six control parameters are summarized in \tref{tab:control_parameters}.

	\begin{table}
		\fl \caption{\label{tab:control_parameters}The six control parameters in the non-dimensionalized hydrodynamic description \eref{eq:continuum_eq_density} -- \eref{eq:continuum_eq_polarization} of the repulsive chiral active matter system.}
		\begin{indented}
			\item[]\begin{tabular}{@{}lll}
				\br
				parameter & definition & description\\
				\mr
				${\rm Pe}_a$ & $v_0^a\,\tau/\ell$ & P\'eclet number\\
				$\mathcal{R}_a$ & $2\,k\,\mu_r\,R_{r}^4\,\pi\,\rho_0^a\,\tau/(3\,\ell^2)$ & strength of the repulsive force\\
				$z$ & $\zeta\,\rho_0\,\tau/\ell$ & particle velocity-reduction due to the neighbors\\
				$D_{\rm t}$ & $\xi\,\tau/\ell^2$ & translational diffusion\\
				$\Omega_a$ & $\omega_a\,\tau$ & intrinsic frequency\\
				$g_{ab}$ & $K_{ab}\,\mu_{\theta}\,R_{\theta}^2\,\pi\,\rho_0^b\,\tau/2$ & orientational coupling parameter\\
				\br
			\end{tabular}
		\end{indented}
	\end{table}

	To put the hydrodynamic equations derived here into the context of earlier continuum models for chiral active particles, we note that for one species of particles without steric repulsion (i.e.~$\mathcal{R}_a=z=0$), our hydrodynamic equations match those given by Liebchen and Levis in \cite{liebchen_2017_collective_behavior_chiral_active_matter_pattern_formation_flocking}. Further, when considering several different, non-mutually coupled species, our equations are in agreement with those derived by Fruchart et al.~\cite{fruchart_2021_non-reciprocal_phase_transitions}. However, Fruchart et al.~neither considered the presence of an intrinsic frequency (i.e.~$\Omega_a=0$) nor volume exclusion. In this work, we specifically allow for non-reciprocal couplings between different species of chiral particles, which do not only (anti-)align but also sterically interact due to the finite particle size.
	
	In the present paper, we study the collective behavior based on the hydrodynamic Equations \eref{eq:continuum_eq_density} -- \eref{eq:continuum_eq_polarization} following essentially two strategies. First, we employ a linear stability analysis starting from the uniform, rotationally isotropic state given by $(\rho^a, \bm{w}^a) = (\rho^a_0, \bm{0})$. This state corresponds, in fact, to the ``trivial'' solution of Equations \eref{eq:continuum_eq_density} -- \eref{eq:continuum_eq_polarization}. Second, we perform numerical simulations of the full hydrodynamic equations in two-dimensional periodic systems. For the numerical simulations, we use a pseudo-spectral code in combination with an operator splitting technique, allowing us to treat the linear operator exactly in the fourth-order Runge Kutta time integration. We choose the initial state to be a slightly perturbed disordered state of zero polarization $\bm{w}(\bm{r},0)=\bm{0}$ and constant density \mbox{$\rho(\bm{r},0)=\rho_0=1$}. The two-dimensional simulation box of size $100\,\ell \times 100\,\ell$ is separated into $256 \times 256$ grid points.

	\section{One species}
	\label{sec:one_species}
	We start by investigating a system with one species ($a=A$) of repulsive chiral active particles, which allows us to focus on the effect and interplay of intrinsic frequency, steric repulsion, and reciprocal orientational couplings. On the basis of this section's results, we will then address the impact of non-reciprocity by studying a corresponding mixture in the subsequent \sref{sec:two_species}.

	\subsection{Linear stability}
	\label{ssec:one_species_linear_stability}
	
	\subsubsection{Methodology}	
	To examine analytically the linear stability of the disordered, uniform state characterized by $(\rho, \bm{w})= (\rho_0, \bm{0})$, we investigate the dynamical behavior of perturbations of the form
	\begin{equation}
	\label{eq:perturbations_one_species}
	\fl \rho'(\bm{r},t) = \int \hat{\rho}(k) \, \rme^{i\bm{k}\cdot\bm{r}+\sigma(k)t} \, {\rm d}\bm{k}, \quad \bm{w}'(\bm{r},t) = \int \hat{\bm{w}}(k) \, \rme^{i\bm{k}\cdot\bm{r}+\sigma(k)t} \, {\rm d}\bm{k}.
	\end{equation}
	As expressed by \Eref{eq:perturbations_one_species}, the system is subject to perturbations involving all wave numbers $k$. We assume these perturbations to be of plane wave form with wave vector $\bm{k}$, (complex) growth rate $\sigma(k)$ and amplitudes $\hat{\rho}(k)$ and $\hat{\bm{w}}(k)$.
	Here, $\sigma$ depends only on the wave number $k=\vert \bm{k} \vert$, because we study the stability of the \textit{isotropic} base state. We now insert the ansatz $\rho(\bm{r},t)=\rho_0+\rho'(\bm{r},t)$, $\bm{w}(\bm{r},t)=\bm{w}'(\bm{r},t)$ into the evolution equations \eref{eq:continuum_eq_density} and \eref{eq:continuum_eq_polarization} (restricting them to one species) and assume $\rho'$ and $\bm{w}'$ to be small. Linearization with respect to the perturbation then leads to a decoupling with respect to $k$. For each $k$, this yields an eigenvalue problem given by
	\begin{equation}
		\label{eq:one_species_linear_dynamics}
		\sigma(k)  \left(\matrix{\hat{\rho}(k) \cr \hat{w}_x(k) \cr \hat{w}_y(k)}\right) = \bm{\mathcal{M}}_1(k) \cdot \left(\matrix{\hat{\rho}(k) \cr \hat{w}_x(k) \cr \hat{w}_y(k)}\right)
	\end{equation}
	with
	\begin{equation}
		\label{eq:one_species_linear_dynamics_matrix}
		\eqalign{
			\fl \bm{\mathcal{M}}_1(k)
			= \left(\matrix{-(\mathcal{R}\,\rho_0 +D_{\rm t})\,k^2  & -i\,v(\rho_0)\,k_x & -i\,v(\rho_0)\,k_y  \cr
				-\frac{i}{2}\,(v(\rho_0)-z\,\rho_0)\, k_x & -\mathcal{D}\,k^2 + g_{AA}\,\rho_0 - 1 & \frac{v^2(\rho_0)}{4\,b} \, \Omega_A \,k^2 - \Omega_A \cr
				-\frac{i}{2}\,(v(\rho_0)-z\,\rho_0)\, k_y & -\frac{v^2(\rho_0)}{4\,b} \, \Omega_A \,k^2 + \Omega_A & -\mathcal{D}\,k^2 + g_{AA}\,\rho_0 - 1
			}\right),
		}
	\end{equation}
	where $\mathcal{D} = v^2(\rho_0)/(2\,b)+ D_{\rm t}$, and effective velocity $v(\rho_0) = {\rm Pe} - z\,\rho_0$. From \Eref{eq:one_species_linear_dynamics}, we can derive analytical expressions for the (complex) growth rates $\sigma(k)$, which play the roles of eigenvalues. Note that the real parts of the three eigenvalues $\sigma(k)$ determine the actual growth in time, whereas the imaginary parts are related to oscillatory behavior. We thus focus mainly on investigating the real parts, ${\rm Re}(\sigma)$.
	In particular, the disordered state is linearly stable only if ${\rm Re}(\sigma(k))<0$ for all $k$. In contrast, it is linearly unstable as soon as ${\rm Re}(\sigma(k))>0$ for any $k$. In our investigation, we monitor all three functions ${\rm Re}(\sigma(k))$. To interpret the behavior, we here assume that the largest value and corresponding eigenvector determine the type of emerging dynamics at short times. This assumption is later checked by means of numerical continuum simulations.
	
	The eigenvalue equation \eref{eq:one_species_linear_dynamics} and, in particular, matrix \eref{eq:one_species_linear_dynamics_matrix} already allow us to deduce some general features of the linear dynamics.
	First, we note that the growth rates at $k=0$ play a special role as they are generally related to the change of the spatially integrated value of the hydrodynamic quantities, corresponding to long-wavelength perturbations. Importantly, such a perturbation can occur only in the polarization (signaling flocking) but not in the particle density. 
	To see this, we recall that $\rho(\bm{r},t)$ is conserved by means of the continuity equation~\eref{eq:continuum_eq_density}, such that
	\begin{equation}
		\frac{\rm d}{{\rm d}t} \int \rho'(\bm{r},t) \, {\rm d}\bm{r} = \frac{\rm d}{{\rm d}t} \, \hat{\rho}(k=0) = 0 .
	\end{equation}
	As seen from \Eref{eq:one_species_linear_dynamics}, this implies that the growth rate of the corresponding mode at $k=0$ must vanish. Hence, our system generally features at least one $\sigma(k=0)=0$. The two remaining eigenvalues at $k=0$ are related to long-wavelength fluctuations of the polarization $\hat{\bm{w}}(k=0)=\int\bm{w}'(\bm{r},t)\,{\rm d}\bm{r}$, whereby ${\rm Re(\sigma(k=0))}>0$ or ${\rm Re(\sigma(k=0))}<0$ indicate that such a fluctuation grows or is suppressed.
	
	Investigating the stability at finite $k$, on the other hand, is equivalent to taking the self-propulsion of swimmers ($v(\rho_0)$) into account. This is because in matrix \eref{eq:one_species_linear_dynamics_matrix}, all terms proportional to $k$ or $k^2$, which stem from gradient terms, are only non-zero for $v(\rho_0)>0$, expect for the diffusional terms ($\sim D_{\rm t}, \mathcal{R}$).
	
	From matrix \eref{eq:one_species_linear_dynamics_matrix} we can further extract the parameters with the most important effects on the linear stability of the system. Specifically, it is seen that the translational diffusion coefficient $D_{\rm t}$ and steric repulsion parameter $\mathcal{R}$ only appear in the diagonal entries and scale with $k^2$, ensuring the stability at large wave numbers. Further, the average density $\rho_0$ only appears as a prefactor of $\mathcal{R}$, $z$, and $g_{AA}$, and thus has no direct effect itself. The P\'eclet number ${\rm Pe}$ only appears in $v(\rho_0)$, such that only the difference between ${\rm Pe}$ and $z\,\rho_0$ has an impact on the dynamics, but not ${\rm Pe}$ itself. Hence, all possible scenarios are governed by parameters $z$, $g_{AA}$, and $\Omega_A$.

	\subsubsection{Results}
	\label{sssec:one_species_results}
	As the analytical expressions for $\sigma(k)$ at arbitrary frequencies $\Omega_A$ are long, high-order polynomials, we here consider the exemplary case of $\Omega_A=0.1$. The question of how the value of $\Omega_A$ affects the dynamics is discussed in \sref{ssec:one_species_effect_intrinsic_frequency}.

\begin{figure}[ht]
		\centering
		\includegraphics[scale=1]{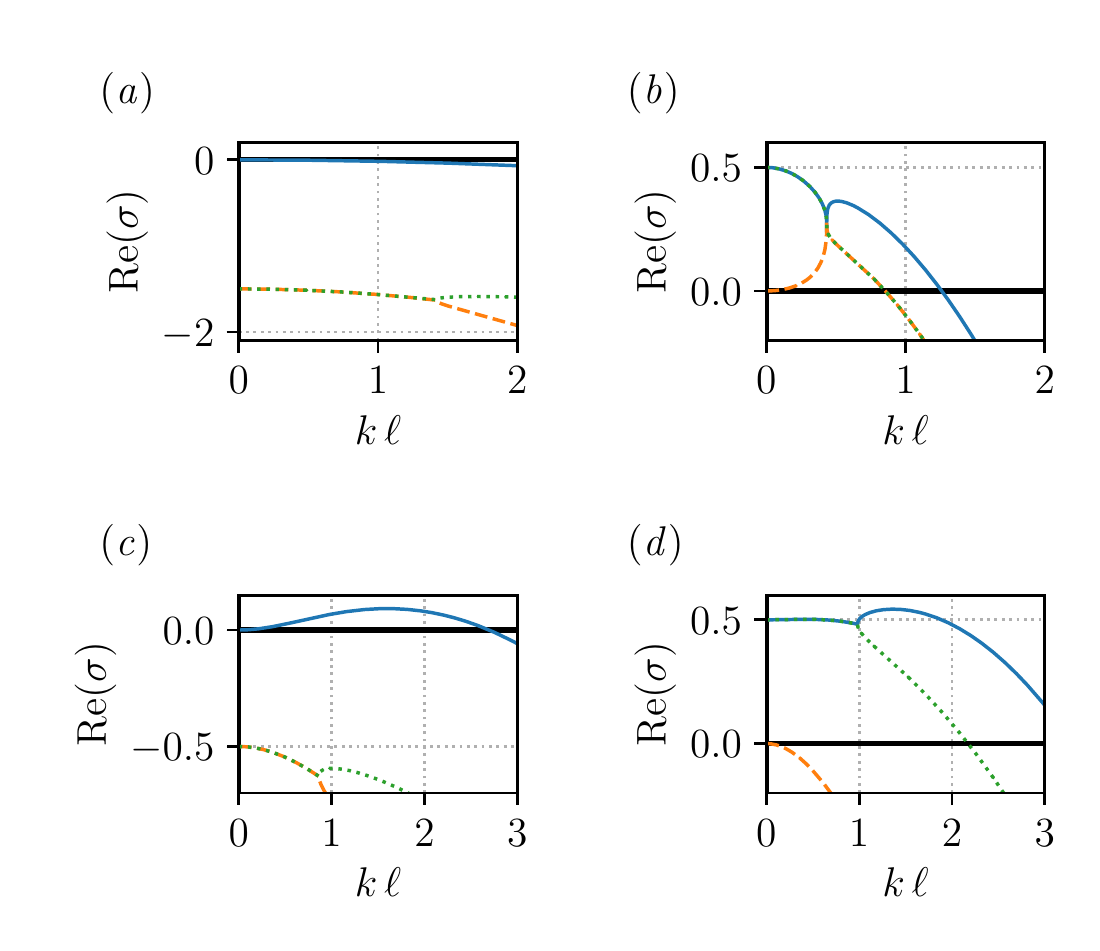}
		\caption{
			Real parts of the growth rates of the four emerging stability scenarios in the one-species system of chiral active particles for different orientational coupling strengths $g_{AA}$ and velocity-reduction parameters $z$. The intrinsic frequency is exemplarily set to $\Omega_A=0.1$. Other parameters are ${\rm Pe}=1.5$, $\mathcal{R}=0.1$, $\rho_0=1$. Different colors indicate the three different growth rates in the one-species system. (\textit{a}) Stable disordered phase for $g_{AA}=-0.5$, $z=0.75\, {\rm Pe/\rho_0}$, and $D_{\rm t}=0.01$. (\textit{b}) Flocking for $g_{AA}=1.5$, $z=1.25 \,{\rm Pe/\rho_0}$, and $D_{\rm t}=0.3$. (\textit{c}) Pure MIPS for $g_{AA}=0.5$, $z=0.75\, {\rm Pe/\rho_0}$, and $D_{\rm t}=0.05$. (\textit{d}) MIPS combined with flocking for $g_{AA}=1.5$, $z=0.75\, {\rm Pe/\rho_0}$, and $D_{\rm t}=0.1$.}
		\label{fig:growthRatesOneSpecies}
	\end{figure}

\begin{figure}[ht]
		\centering \includegraphics[width=0.7\textwidth]{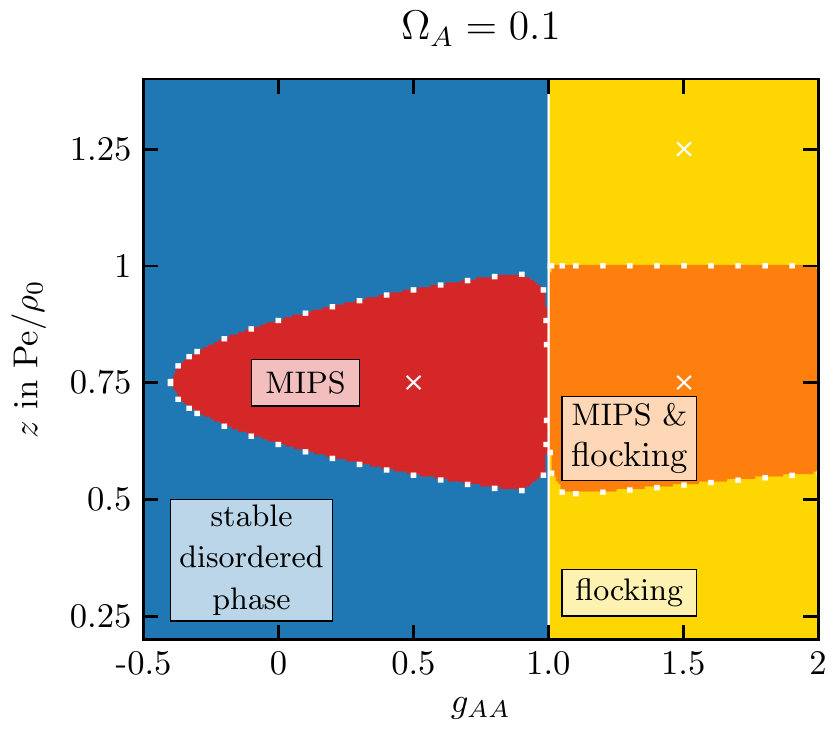}
		\caption{Stability diagram obtained for the one-species system of repulsive chiral active particles with alignment interactions. We exemplarily show the result for intrinsic frequency $\Omega_A=0.1$. Depending on the orientational coupling strength $g_{AA}$ and velocity-reduction parameter $z$, the system can exhibit different types of instabilities. The disordered state (blue) is stable for weak alignment, $g_{AA}\,\rho_0<1$, or anti-alignment, $g_{AA}<0$. Increasing the alignment between particles ($g_{AA}\,\rho_0>1$), the system undergoes a flocking transition (yellow). For specific values of $z$, the system exhibits MIPS -- either pure for weak alignment (red) or in combination with flocking for strong alignment (orange). Analytically estimated stability conditions for $z$ are plotted as white squares. The white crosses correspond to parameters used in the numerical continuum simulations in \sref{ssec:one_species_numerical_simulations}. Other parameters are ${\rm Pe}=1.5$, $\mathcal{R}=0.1$, $D_{\rm t}=0.01$, and $\rho_0=1$.}
		\label{fig:phaseDiagramOneSpecies}
	\end{figure}

	The remaining relevant parameters are $z$ and $g_{AA}$, whereby $g_{AA}>0$ favors alignment and $g_{AA}<0$ anti-alignment. Investigating the behavior of the eigenvalues and eigenvectors as functions of these parameters, we observe essentially four types of scenarios. These correspond to the stable disordered state, a flocking state, MIPS or MIPS in combination with flocking. The four scenarios show distinct behavior of the real parts of the growth rates. The main characteristics are summarized in \tref{tab:one-species_eigenvalues_order_parameters}. The table also provides information about the magnitude of the polarization field and the shape of the density distribution characterizing each scenario. These quantities are obtained within the continuum simulations as described in \sref{ssec:one_species_numerical_simulations}. Examples for the $k$-dependence of the eigenvalues for each scenario are shown in \fref{fig:growthRatesOneSpecies}. In addition, we provide in \fref{fig:phaseDiagramOneSpecies} an (analytically determined) stability diagram showing which type of scenario occurs at a given value of $z$ and $g_{AA}$.

	\begin{table}
		\fl \caption{\label{tab:one-species_eigenvalues_order_parameters}Real parts of growth rates (eigenvalues) of the stability matrix \eref{eq:one_species_linear_dynamics_matrix} with corresponding order parameters (absolute value of polarization, $\vert \bm{w}(\bm{r},t) \vert$, and probability distribution of density, $p(\rho(\bm{r},t)$), obtained after the initial transient regime) for the four stability scenarios observed in the one-species system. The eigenvectors \mbox{$\bm{v}(k=0)=(\hat{\rho}, \hat{w}_x, \hat{w}_y)^{\rm T}$} corresponding to eigenvalues $\sigma_{1/2/3}$ are \mbox{$\bm{v}_1(k=0)=(1,0,0)^{\rm T}$} and \mbox{$\bm{v}_{2/3}(k=0)=(0,\hat{w}_{x_{1,2}},\hat{w}_{y_{1,2}})^{\rm T}$} with $\hat{w}_{x_{1,2}},\hat{w}_{y_{1,2}} \in \mathbb{C}$.}
		\begin{indented}
			\item[]\begin{tabular}{p{1.2cm} p{4.4cm} p{4.2cm}}
				\br
				 & real part of eigenvalues, ${\rm Re}(\sigma(k))$ & \parbox{4.2cm}{order parameters\\ $\forall \ \bm{r},t$ after transient regime}\\
				\br
				disordered & \parbox{4.4cm}{${\rm Re}(\sigma_1(k=0))=0$ \\ ${\rm Re}(\sigma_{2/3}(k=0))<0$\\ ${\rm Re}(\sigma_{1/2/3}(k>0))<0$}   & \parbox{4.2cm}{$\vert \bm{w}(\bm{r},t) \vert = 0$ \\ $p(\rho)$ unimodal ($\rho(\bm{r},t)={\rm const.}$) }\\
				\mr
				flocking & \parbox{4.4cm}{${\rm Re}(\sigma_1(k=0))=0$ \\ ${\rm Re}(\sigma_{2/3}(k=0))>0$\\ maximum of ${\rm Re}(\sigma_{2/3}(k))$ at $k=0$} & \parbox{4.2cm}{$\vert \bm{w}(\bm{r},t) \vert >0$ with $\vert \langle \bm{w}(t) \rangle \vert >0 $\\ $p(\rho)$ unimodal ($\rho(\bm{r},t)={\rm const.}$)}\\
				\mr
				MIPS & \parbox{4.4cm}{${\rm Re}(\sigma_1(k=0))=0$ \\ ${\rm Re}(\sigma_{2/3}(k=0))<0$\\ maximum of ${\rm Re}(\sigma_1(k))$ at $k>0$} & \parbox{4cm}{$\vert \bm{w}(\bm{r},t) \vert =0$ \\ $p(\rho)$ bimodal ($\rho(\bm{r},t)\neq{\rm const.}$)}\\
				\mr
				\parbox{1.2cm}{MIPS \&\\ flocking} & \parbox{4.4cm}{${\rm Re}(\sigma_1(k=0))=0$ \\ ${\rm Re}(\sigma_{2/3}(k=0))>0$\\ maximum of ${\rm Re}(\sigma_{2/3}(k))$ at $k>0$} & \parbox{4cm}{$\vert \bm{w}(\bm{r},t) \vert >0$ within clusters\\ $p(\rho)$ bimodal ($\rho(\bm{r},t)\neq{\rm const.}$)}\\
				\br
			\end{tabular}
		\end{indented}
	\end{table}

	\Fref{fig:growthRatesOneSpecies}(\textit{a}) illustrates the behavior of the eigenvalues within the disordered state $(\rho_0, \bm{0})$ (blue region in \fref{fig:phaseDiagramOneSpecies}). Here, ${\rm Re}(\sigma(k))\leq 0$ for all $k$, reflecting that perturbations of any type and at all wave numbers decrease in time. As soon as ${\rm Re}(\sigma)$ becomes positive, collective dynamics start to emerge. 
	
	We first look at the emergence of a flocking phase (see \fref{fig:growthRatesOneSpecies}(\textit{b}) and yellow region in \fref{fig:phaseDiagramOneSpecies}). Flocking is generally characterized by the emergence of a large-scale, non-zero polarization, that is, $ \vert \langle\bm{w}\rangle \vert \neq 0 $, where $\langle \cdot\rangle$ denotes the spatial average. We recall that, in an active fluid, such a polarization implies ordered motion of the particles. To find the corresponding conditions, we consider the complex growth rates at $k=0$, 
	\begin{subnumcases} {\label{eq:one_species_flocking}}
		\sigma_{1}(k=0) = 0\\
		\sigma_{2/3}(k=0) = g_{AA}\,\rho_0-1 \pm i\,\Omega_A \label{eq:one_species_flocking_polarization} .
	\end{subnumcases}
	The corresponding eigenvectors \mbox{$\bm{v}(k=0)=(\hat{\rho}_0, \hat{w}_{x,0}, \hat{w}_{y,0})^{\rm T}$} are \mbox{$\bm{v}_1(k=0)=(1,0,0)^{\rm T}$} and \mbox{$\bm{v}_{2/3}(k=0)=(0,\hat{w}_{x_{1,2}},\hat{w}_{y_{1,2}})^{\rm T}$}, where $\hat{w}_{x_{1,2}},\hat{w}_{y_{1,2}} \in \mathbb{C}$.
	Hence, the eigenvalue $\sigma_1(k=0)$ reflects the conservation of the particle density, whereas $\sigma_{2/3}(k=0)$ indicate the change of the overall polarization. As seen from \eref{eq:one_species_flocking_polarization}, a flocking instability can only occur for strong alignment couplings characterized by $g_{AA}\,\rho_0>1$.
	Furthermore, while the real parts of growth rates~\eref{eq:one_species_flocking} are independent of the intrinsic frequency $\Omega_A$, the imaginary parts are not. Thus, $\Omega_A$ does not affect the flocking instability itself, but makes the instability oscillatory. Indeed, as shown later in numerical simulations (\sref{ssec:one_species_numerical_simulations}), the flocking phase is time-dependent in the sense that $\bm{w}(\bm{r},t)$ rotates in time. Microscopically, this means that the chiral active particles synchronize, such that they rotate coherently with a frequency, which is not necessarily the same as the intrinsic one. Within the flocking regime, the full real growth rates look like shown in \fref{fig:growthRatesOneSpecies}(\textit{b}) with two ${\rm Re}(\sigma(k=0))>0$, which decrease for increasing $k$. Since this type of instability is not affected by volume exclusion effects (modeled by parameters $\mathcal{R}$ and $z$), Liebchen and Levis \cite{liebchen_2017_collective_behavior_chiral_active_matter_pattern_formation_flocking} observed the same flocking instability \eref{eq:one_species_flocking_polarization} for non-repulsive chiral systems.
	
	Taking volume exclusion into account, the second type of instability is related to (pure) MIPS, where the system separates into high- and low-density regions (see red regions in \fref{fig:phaseDiagramOneSpecies}). The real growth rates typical for pure MIPS are shown in \fref{fig:growthRatesOneSpecies}(\textit{c}). The two eigenvalues with ${\rm Re}(\sigma(k=0))<0$ indicate that fluctuations of the overall polarization are suppressed. From \eref{eq:one_species_flocking_polarization} it then follows that pure MIPS can only occur when the alignment coupling is weak (i.e.~$g_{AA}\,\rho_0<1$) or particles anti-align ($g_{AA}<0$). We further see that the eigenvalue related to density fluctuations increases towards positive values at small $k$ and exhibits a maximum at finite $k>0$. Such a maximum indicates a characteristic length scale of emerging patterns at small times.
	Importantly, we observe this type of instability only in a certain range of velocity-reduction parameters~$z$. In fact, in case of non-chiral active particles, we can analytically compute a condition for $z$ as outlined in \ref{sec:non-chiral_system_MIPS}. For chiral active particles, we use the same approach (that is expanding the eigenvalues up to order$~k^2$) to estimate values of $z$ for which one eigenvalue becomes positive. This yields the white squares in \fref{fig:phaseDiagramOneSpecies}. A more detailed discussion of the role of $\Omega_A$ on MIPS is postponed to \sref{ssec:one_species_effect_intrinsic_frequency}. Here, we first address the effect of increasing alignment interactions $g_{AA}$.
	
	As mentioned before, pure MIPS can only occur when $g_{AA}\,\rho_0<1$ and a violation of this condition yields flocking.
	Thus, a suitable combination of the velocity-reduction parameter~$z$ and alignment can lead to the simultaneous emergence of MIPS and flocking (see orange regions in \fref{fig:phaseDiagramOneSpecies}). The corresponding real growth rates, shown in \fref{fig:growthRatesOneSpecies}(\textit{d}), feature characteristics of both transitions: A positive value of two ${\rm Re}(\sigma(k))$ at $k=0$ (flocking) and a maximum at finite $k>0$ (MIPS).
	
	We note that a similar phenomenon, namely the emergence of clusters consisting of synchronized particles, has also been observed in the absence of steric repulsion at sufficiently large frequencies \cite{liebchen_2017_collective_behavior_chiral_active_matter_pattern_formation_flocking}. However, this so-called ``microflock instability'' is not related to the simultaneous emergence of flocking and MIPS reported in the present paper. Different to \cite{liebchen_2017_collective_behavior_chiral_active_matter_pattern_formation_flocking}, we here consider steric repulsion between particles, leading to density-dependent velocity reduction and, consequently, MIPS. In contrast, the clustering in \cite{liebchen_2017_collective_behavior_chiral_active_matter_pattern_formation_flocking} appears as a ``secondary'' instability not of the isotropic state (considered here) but of the flocking state.

	\subsection{Numerical continuum simulations}
	\label{ssec:one_species_numerical_simulations}
	While the linear stability analysis is a convenient tool to analyze fluctuations around the base state considered, we can neither use it to deduce the dynamics at long times nor the exact form of emerging patterns. In principle, these questions could be solved by performing particle-based simulations of the microscopic Langevin equations \eref{eq:Langevin_r} and \eref{eq:Langevin_theta}. However, extensive parameter scans and long simulation times would make a quick exploration of parameter spaces rather difficult.
	We therefore complement the previous analysis with numerical simulations of the full, non-linear hydrodynamic equations \eref{eq:continuum_eq_density} -- \eref{eq:continuum_eq_polarization}. Specifically, we consider the three parameter sets indicated by the white crosses in \fref{fig:phaseDiagramOneSpecies}. This choice corresponds to parameters which lie well within the predicted stability regions of MIPS, flocking, and MIPS combined with flocking, respectively. 
	
	The emerging particle density and polarization density fields are time-dependent. In the following, we focus on times after the initial transient regimes ($t>t_{\rm t}$), whereby the time $t_{\rm t}$, after which the respective system has passed the transient regime, depends on the parameters. The subsequently shown snapshots represent instantaneous fields at $t>t_{\rm t}$. For an illustration of the actual time dependence, we provide videos of the different phases in the supplemental material.
	
	\begin{figure}[h]
		\centering \includegraphics[scale=0.8]{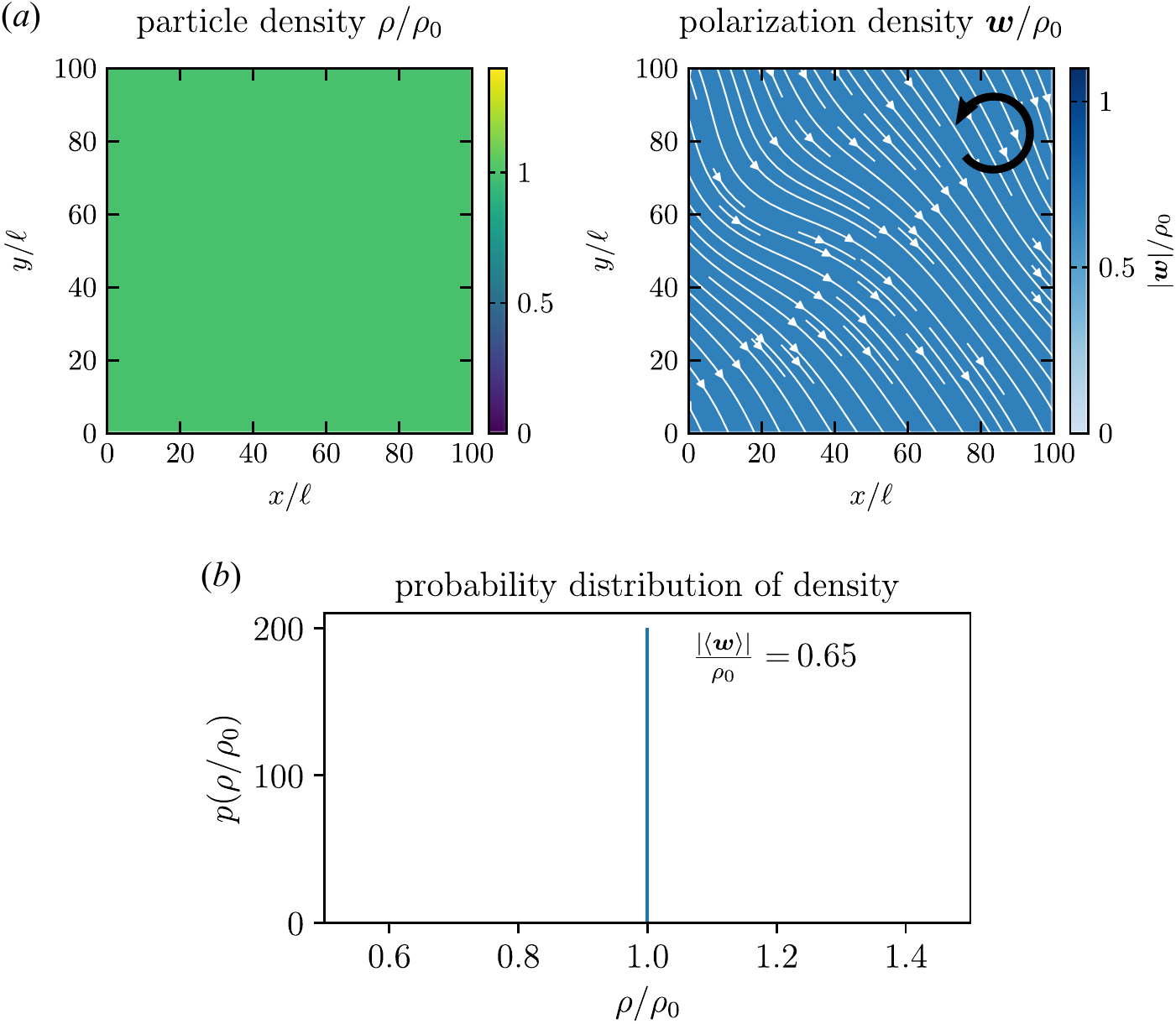}
		\caption{Numerical simulation results of a one-species system undergoing a flocking transition after the initial transient regime, $t>t_{\rm t}$. (\textit{a}) Snapshots of the constant particle density $\rho(\bm{r},t)$ and time-dependent polarization density field $\bm{w}(\bm{r},t)$. White arrows indicate the instantaneous, local direction of $\bm{w}$, the black arrow indicates the rotation of the flock direction in time. 
		(\textit{b}) Time-independent probability distribution $p(\rho)$ and absolute value of spatially-averaged polarization $\vert \langle \bm{w} \rangle \vert$. 
		The parameters are $g_{AA}=1.5$, $z=1.25\,{\rm Pe}/\rho_0$, $\Omega_A=0.1$, ${\rm Pe}=1.5$, $D_{\rm t}=0.3$, $\rho_0=1$, and $\mathcal{R}=0.1$.}
		\label{fig:one_species_numerical_results_flocking}
	\end{figure} 
	
	We start by discussing the flocking phase. Representative snapshots of the spatially resolved particle density $\rho(\bm{r},t)$ and the corresponding time-dependent polarization density field $\bm{w}(\bm{r},t)$ are shown in \fref{fig:one_species_numerical_results_flocking}(\textit{a}).  \Fref{fig:one_species_numerical_results_flocking}(\textit{b}) shows the probability distribution $p(\rho)$ of the local density. As expected from our stability analysis, within the (pure) flocking state, $\rho(\bm{r},t)$ is constant, as indicated by the single sharp peak in $p(\rho)$. 
	The spatially-averaged polarization density field rotates in time and can be described by
	\begin{equation}
		\label{eq:spatially_averaged_polarization}
		\langle \bm{w}(t) \rangle = \vert \langle \bm{w} \rangle \vert \, \rme^{i\,\Omega_{\rm flock}\,t}.
	\end{equation}
	Its amplitude is given by the (time-independent) absolute value
	\begin{equation}
		\label{eq:magnitude_spatially_averaged_polarization}
		\vert \langle \bm{w} \rangle \vert = \bigg\vert \int \bm{w}(\bm{r},t) \,\rmd \bm{r} \bigg\vert \quad \forall \quad t>t_{\rm t},
	\end{equation}
	measuring the ordering of particles and, hence, the ``strength'' of the flock formation. Further, $\Omega_{\rm flock} = \langle \Omega(\bm{r},t) \rangle$ is the spatially-averaged rotation frequency of the flock.
	In the flocking phase, $\vert \langle \bm{w} \rangle \vert$ is non-zero ($\vert \langle \bm{w} \rangle \vert/\rho_0 \approx 0.65$), reflecting ordered motion induced by sufficiently strong local alignment of particles (compare \eref{eq:one_species_flocking_polarization}). These observations comply with the linear stability analysis (see \fref{fig:growthRatesOneSpecies}(\textit{b})). The stability analysis further predicts that the emerging state is time-dependent since ${\rm Im}(\sigma(k=0))\neq 0$ (see \Eref{eq:one_species_flocking_polarization}), whereby the imaginary part is given by the intrinsic frequency $\Omega_A$. Indeed, we find in our numerical simulation that the direction of polarization rotates in time in counter clockwise direction with rotation frequency $\Omega_{\rm flock}=0.075$. Since $\Omega_{\rm flock}$ is the spatially-averaged rotation frequency with variance of $O(10^{-7})$ and stays constant after a transient regime ($t>t_{\rm t}$), we deduce that the entire flock rotates with the same frequency $\Omega_{\rm flock}$. This flock rotation frequency is somewhat smaller than the intrinsic frequency of each particle, $\Omega_A=0.1$, which matches very well the prediction of linear stability analyses around a flocking state in reciprocal one-species chiral systems in \cite{liebchen_2017_collective_behavior_chiral_active_matter_pattern_formation_flocking}. Microscopically, the non-zero value of $\Omega_{\rm flock}$ implies that particles synchronize and thereby rotate in a coherent manner.
	For non-chiral active particles ($\Omega_A=0$), numerical simulations show that the flock retains its direction of motion, as one would expect.

	\begin{figure}[h]
		\centering \includegraphics[scale=0.8]{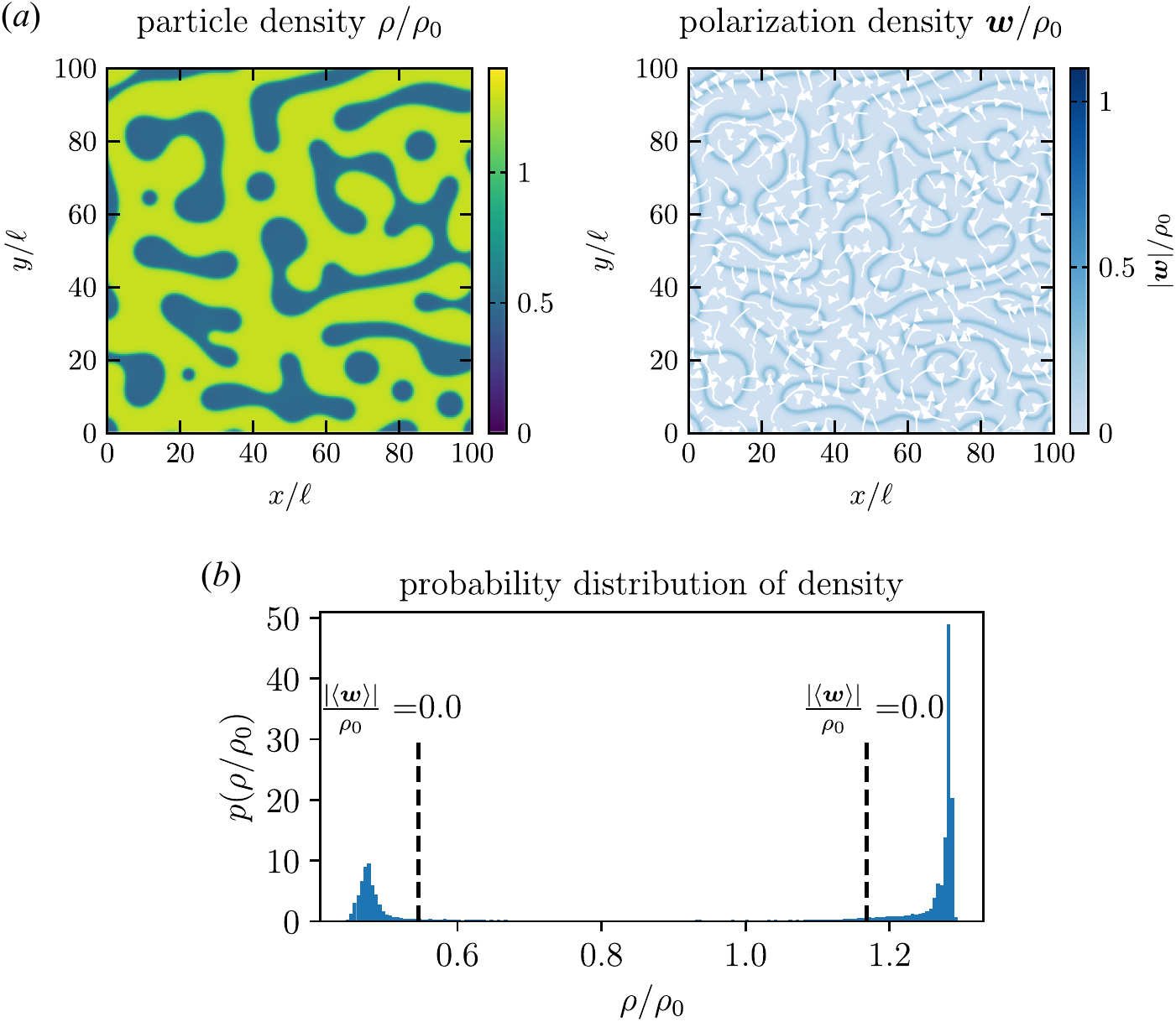}
		\caption{Numerical simulation results of MIPS in a one-species system. (\textit{a}) Snapshots of the time-dependent particle density $\rho(\bm{r},t)$ and polarization density field $\bm{w}(\bm{r},t)$. The clusters of enhanced density grow in time. White arrows indicate the instantaneous, local direction of $\bm{w}$. (\textit{b}) Instantaneous probability distribution $p(\rho)$ and absolute value of spatially-averaged polarization $\vert \langle \bm{w} \rangle \vert$ in regions of low densities (smaller than left dashed vertical line) and large densities (larger than right dashed vertical line). 
		The parameters are $g_{AA}=0.5$, $z=0.75\,{\rm Pe}/\rho_0$, $\Omega_A=0.1$, ${\rm Pe}=1.5$, $D_{\rm t}=0.05$, $\rho_0=1$, and $\mathcal{R}=0.1$.}
		\label{fig:one_species_numerical_results_MIPS}
	\end{figure}

	We now turn to (pure) MIPS. In this case, numerical simulations of the full hydrodynamic equations yield particle and polarization density fields as shown in the snapshots in \fref{fig:one_species_numerical_results_MIPS}(\textit{a}). From $\rho(\bm{r},t)$, we observe the emergence and growth of clusters with local density larger than $\rho_0$. In fact, the cluster formation is a consequence of self-trapping mechanisms due to a reduction of self-propulsion velocity in crowded situations (e.g., \cite{cates_2015_MIPS_review,elena_2021_phase_separation_self-propelled_disks}). As a result of the cluster formation, $p(\rho)$ has two peaks: one at low density and the other one at larger density (\fref{fig:one_species_numerical_results_MIPS}(\textit{b})). We also find that the polarization \textit{within} the clusters vanishes, $\vert \bm{w}(\bm{r},t) \vert =0 \ \forall \ \bm{r},t$. These observations are in line with the linear stability analysis (see \fref{fig:growthRatesOneSpecies}(\textit{c})), which predicts a suppression of  overall polarization and a characteristic length scale of (density) patterns at short times. However, the stability analysis cannot predict the slow but steady growth of the clusters seen in the simulations as time proceeds. Neither it can predict the small, non-zero polarization of the cluster interfaces.
	
	\begin{figure}[h]
		\centering \includegraphics[scale=0.8]{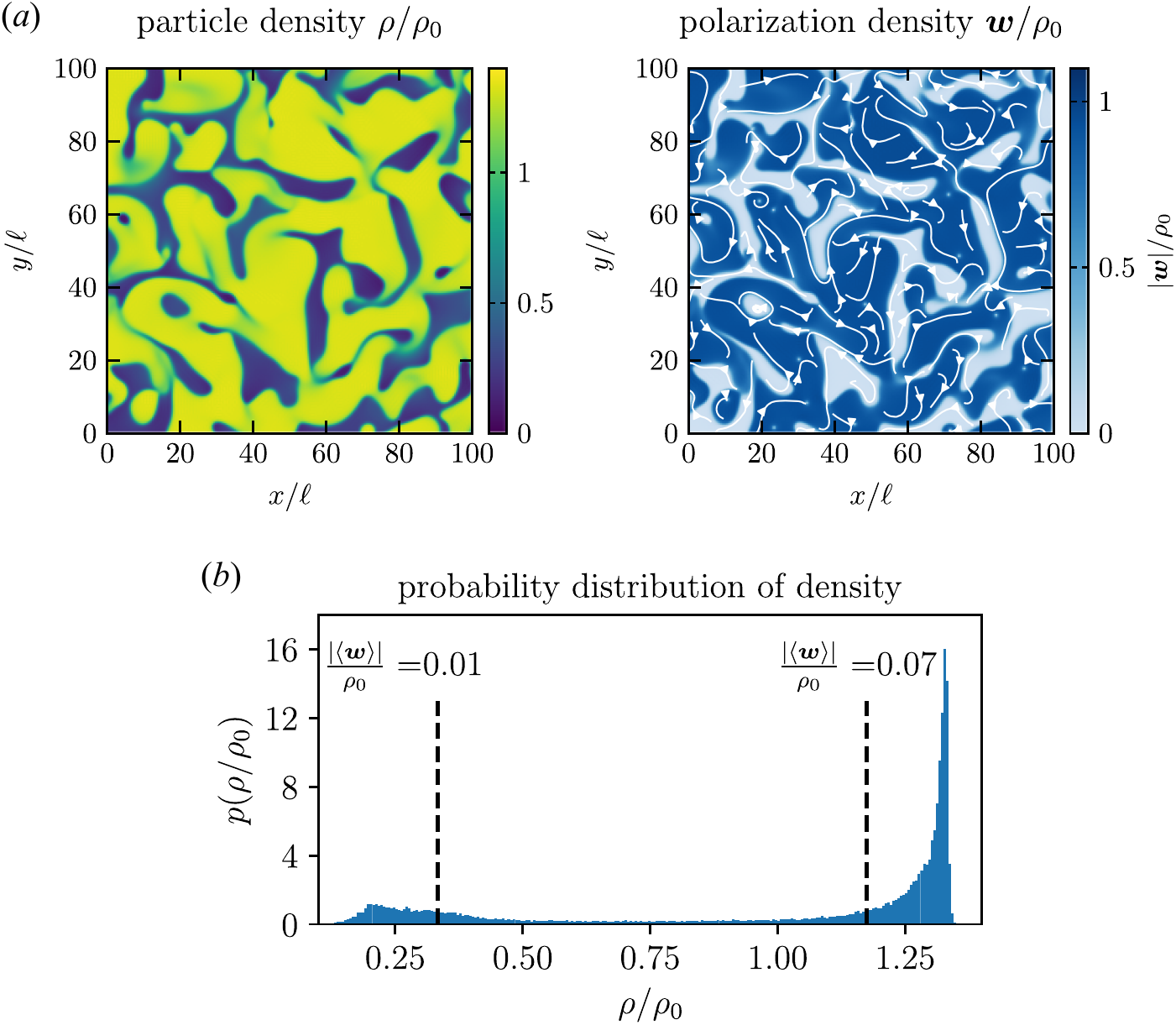}
		\caption{Numerical simulation results of MIPS combined with flocking in a one-species system after the initial transient regime, $t>t_{\rm t}$. (\textit{a}) Representative snapshots of the time-dependent particle density $\rho(\bm{r},t)$ and polarization density field $\bm{w}(\bm{r},t)$. White arrows indicate the instantaneous, local direction of $\bm{w}$. (\textit{b}) Time-independent probability distribution $p(\rho)$ and absolute value of spatially-averaged polarization $\vert \langle \bm{w} \rangle \vert$ in regions of low densities (smaller than left dashed vertical line) and large densities (larger than right dashed vertical line).
		The parameters are $g_{AA}=1.5$, $z=0.75\,{\rm Pe}/\rho_0$, $\Omega_A=0.1$, ${\rm Pe}=1.5$, $D_{\rm t}=0.1$, $\rho_0=1$, and $\mathcal{R}=0.1$.}
		\label{fig:one_species_numerical_results_MIPS_flocking}
	\end{figure}

	Finally, we consider the case of MIPS combined with flocking. As seen from the snapshots in \fref{fig:one_species_numerical_results_MIPS_flocking}(\textit{a}), this case is characterized by cluster formation in $\rho(\bm{r},t)$, where the entire clusters now display a non-vanishing polarization $\vert \bm{w} \vert$. As in the case of pure MIPS, $p(\rho)$ has two peaks, though with non-vanishing $\vert \langle \bm{w}\rangle \vert$ in the high-density regions (see \fref{fig:one_species_numerical_results_MIPS_flocking}(\textit{b})). Again these observations conform with the linear stability analysis (\fref{fig:growthRatesOneSpecies}(\textit{d})). Besides the non-vanishing polarization within clusters, a further difference between pure MIPS and MIPS combined with flocking becomes apparent when looking at the time-evolution of the density fields (see supplemental videos). Different to the steady growth of clusters obtained for pure MIPS, the combined MIPS and flocking case is characterized by rather constant cluster sizes. The existing clusters merge and break-up constantly, while particles within the clusters form short-living flocks. Describing each of these flocks by \Eref{eq:spatially_averaged_polarization}, we can numerically extract the spatially-averaged rotation frequency $\Omega_{\rm flock} = 0.037$ with variance ${\rm var}(\Omega)=0.004$. The overall behavior is reminiscent of the ``interrupted motility-induced phase separation'' observed in experiments, which show that alignment of active Janus colloids interrupts the phase separation process, eventually leading to a fluctuating but non-increasing average cluster size \cite{van_der_linden_2019_interrupted_mips}.

	Taken altogether, the results of our numerical simulations of the full, non-linear hydrodynamic equations are consistent with the predictions of the stability analysis regarding the type of emerging collective behavior. This holds even when both order parameters are involved, as in the case for the combined MIPS and flocking instability.

	\subsection{Effect of intrinsic frequency}
	\label{ssec:one_species_effect_intrinsic_frequency}
\begin{figure}[htbp]
		\centering
		\includegraphics[scale=1]{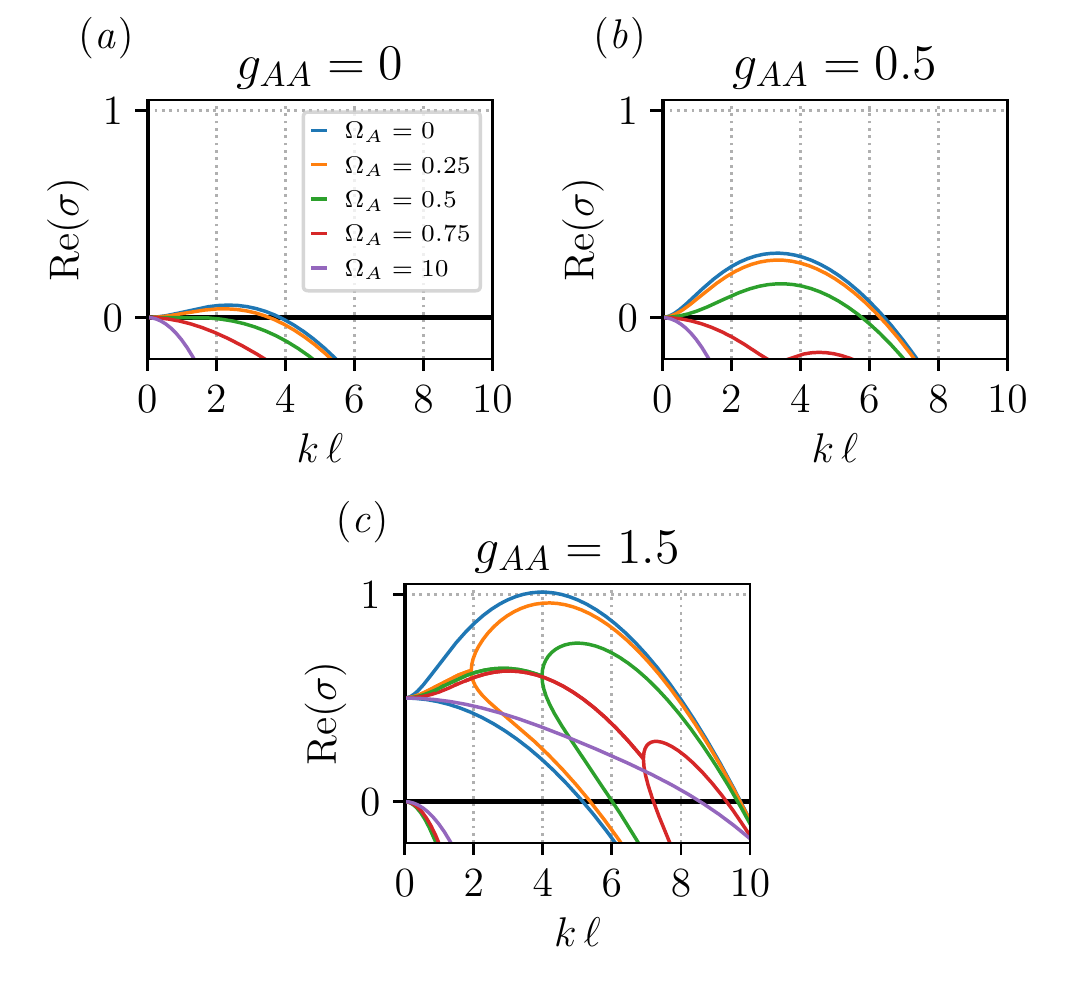}
		\caption{Effect of intrinsic frequency $\Omega_A$ on MIPS for different alignment strengths $g_{AA}$. The relevant real growth rate obtained from the linear stability analysis \eref{eq:one_species_linear_dynamics} is shown for the case of (\textit{a}) no alignment, (\textit{b}) weak alignment, and (\textit{c}) strong alignment. The exemplary parameters chosen for these plots are ${\rm Pe=1.5}$, $\mathcal{R}=0.1$, $z=0.75\,{\rm Pe}/\rho_0$, $D_{\rm t}=0.01$, and $\rho_0=1$.}
		\label{fig:effect_intrinsic_frequency}
	\end{figure}

	We now come back to the question of how the intrinsic frequency $\Omega_A$ of the chiral active particles affects MIPS. So far, this has been studied by particle-based simulations but not via a hydrodynamic theory \cite{Liao_2018_circle_swimmers_monolayer}. Here, we investigate this question on the basis of a linear stability analysis. Results for the eigenvalues as functions of $k$ are presented in \fref{fig:effect_intrinsic_frequency}, where we consider three values of alignment coupling.
	
	In systems without alignment interactions ($g_{AA}=0$), the instability related to (pure) MIPS remains for small intrinsic frequencies $\Omega_A \neq 0.25$ (\fref{fig:effect_intrinsic_frequency}(\textit{a})). In this range, the (relevant) growth rate ${\rm Re}(\sigma(k))$ is zero at $k=0$ and has a maximum at finite wave number. However, when the particles rotate with a larger frequency ($\Omega_A=0.5,\, 0.75,\, 10$), the maximum disappears and the disordered phase becomes stable. Thus, we conclude that the intrinsic frequency of chiral particles generally opposes MIPS. This prediction is consistent with results from particle-based simulations of non-aligning chiral active particles by Liao et al.~\cite{Liao_2018_circle_swimmers_monolayer}, who showed that MIPS only occurs for small intrinsic frequencies.
	
	Switching on the alignment coupling between the particles, but keeping it weak (e.g.~$g_{AA}=0.5$), the linear stability analysis still predicts pure MIPS for slowly rotating active particles (\fref{fig:effect_intrinsic_frequency}(\textit{b})). In fact, the range of $\Omega_A$ where MIPS occurs is even extended relative to the case $g_{AA}=0$, as seen from the curve pertained to $\Omega_A=0.5$ in \fref{fig:effect_intrinsic_frequency}(\textit{b}). 
	
	Further increasing the alignment strength to $g_{AA}=1.5$, the system undergoes a flocking transition, characterized by ${\rm Re}(\sigma(k=0))>0$ for all $\Omega_A$ considered. As noted in \sref{ssec:one_species_linear_stability}, the very appearance of a flocking transition is independent from the intrinsic frequency of the particles (see \Eref{eq:one_species_flocking_polarization}). In contrast, whether or not MIPS occurs as an additional feature depends on the intrinsic frequency and only happens when particles rotate slowly enough. Interestingly, we see from \fref{fig:effect_intrinsic_frequency}(\textit{c}) that strong alignment allows for MIPS in systems with intrinsic frequencies as large as $\Omega_A=0.75$. Only when the active particles rotate even faster, e.g. $\Omega_A=10$, MIPS is suppressed.
	
	These observations from the linear stability analysis, in particular, the suppression of MIPS for large intrinsic frequencies and the promotion of MIPS due to particle alignment, are also observed in our simulations of the full, non-linear hydrodynamic equations \mbox{\eref{eq:continuum_eq_density} -- \eref{eq:continuum_eq_polarization}} (not shown). However, due to numerical instabilities we cannot consider the same parameters as chosen in \fref{fig:effect_intrinsic_frequency}.

	\section{Two species}
	\label{sec:two_species}
	We now turn to a binary system ($a=A,B$) of chiral active particles, focusing on the impact of \textit{non-reciprocal} alignment couplings. To this end, we employ again the previously established combination of linear stability analyses and numerical solutions of the full continuum equations \eref{eq:continuum_eq_density} -- \eref{eq:continuum_eq_polarization}. Clearly, the two-species system involves a large set of parameters, making a \textit{complete} investigation of the full parameter space a rather overwhelming task. Here we therefore focus on some representative parameter combinations. We recall in this context that several parameters characterizing the two species individually have already set equal (this concerns the steric repulsion, the self-propulsion velocity, and diffusion). The main control parameters for the two-species system are therefore the alignment coupling parameters $g_{ab}$ and the intrinsic frequencies $\Omega_a$. Using the results obtained in \sref{sec:one_species} for the one-species system as a reference, our main goal is to explore the impact of non-reciprocity.
	
	\subsection{Linear stability analysis}
	\label{ssec:two_species_linear_stability}
	In a binary system, the trivial solution to the hydrodynamic equations \eref{eq:continuum_eq_density} -- \eref{eq:continuum_eq_polarization} is given (as in the one-species system) by a disordered state with zero polarization $\bm{w}^A(\bm{r},t)=\bm{w}^B(\bm{r},t)=\bm{0}$ and constant density $\rho^A(\bm{r},t)=\rho_0^A$, $\rho^B(\bm{r},t)=\rho_0^B$.
	Assuming again that the perturbations scale with $\sim\rme^{\sigma(k)t}$ (compare \Eref{eq:perturbations_one_species}), and linearizing, we obtain the eigenvalue equation
	\begin{equation}
	\label{eq:two_species_linear_dynamics}
		\sigma(k)\,\bm{v}(k) = \bm{\mathcal{M}}_2(k) \cdot \bm{v}(k).
	\end{equation}
	In \Eref{eq:two_species_linear_dynamics}, the eigenvector \mbox{$\bm{v} = (\hat{\rho}^A, \hat{w}_x^A, \hat{w}_y^A, \hat{\rho}^B, \hat{w}_x^B, \hat{w}_y^B)^{\rm T}$} is now six-dimensional, containing the possible perturbations of the particle densities and the two components of the polarizations for each species. Further, the $6\times 6$ matrix $\bm{\mathcal{M}}_{2}(k)$ can be written as a combination of two types of submatrices,
	\begin{equation}
			\label{eq:two_species_linear_dynamics_matrix}
			\eqalign{
				\bm{\mathcal{M}}_2(k)
				= \left(\matrix{\bm{P}_A(k)  & \bm{M}_{AB}(k) \cr
					\bm{M}_{BA}(k) & \bm{P}_B(k)
				}\right) .
			}
	\end{equation}
	Here, $\bm{P}_a(k)$ involves the couplings within each species $a=A,B$
	\begin{equation}
		\label{eq:two_species_linear_dynamics_pure_matrix}
		\eqalign{
			\fl \bm{P}_a(k)
			= \left(\matrix{-(\mathcal{R}\,\rho_0^a +D_{\rm t})\,k^2  & -i\,v_a(\rho_0)\,k_x & -i\,v_a(\rho_0)\,k_y  \cr
				-\frac{i}{2}\,(v_a(\rho_0)-z\,\rho_0^a)\, k_x & -\mathcal{D}_a\,k^2 + g_{aa}\,\rho_0 - 1 & \frac{v_a^2(\rho_0)}{4\,b} \, \Omega_a \,k^2 - \Omega_a \cr
				-\frac{i}{2}\,(v_a(\rho_0)-z\,\rho_0^a)\, k_y & -\frac{v_a^2(\rho_0)}{4\,b} \, \Omega_a \,k^2 + \Omega_a & -\mathcal{D}_a\,k^2 + g_{aa}\,\rho_0 - 1
			}\right),
		}
	\end{equation}
	with $\mathcal{D} = v_a^2(\rho_0)/(2\,b)+ D_{\rm t}$ and effective velocity $v_a(\rho_0) = {\rm Pe}_a - z\,(\rho_0^a+\rho_0^b)$.
	The other submatrix involves the couplings between the two species ($ab=AB,BA$)
	\begin{equation}
		\label{eq:two_species_linear_dynamics_mixed_matrix}
		\eqalign{
			\bm{M}_{ab}(k)
			= \left(\matrix{-\mathcal{R}\,\rho_0^a\,k^2  & 0 & 0  \cr
				\frac{i}{2}\,z\,\rho_0^a\, k_x & g_{ab}\,\rho^a_0 & 0 \cr
				\frac{i}{2}\,z\,\rho_0^a\, k_y & 0 & g_{ab}\,\rho^a_0 
			}\right) .
		}
	\end{equation}
	Note that, in case of non-reciprocal alignment couplings, $g_{AB}\neq g_{BA}$, and different intrinsic frequencies, $\Omega_A \neq \Omega_B$, the entire matrix $\bm{\mathcal{M}}_2(k)$ becomes non-symmetric.

	We assume (as in the one-species case) that the six resulting eigenvalues indicate the instability in the two-species system.
	At $k=0$, the growth rates are given by
	\begin{subnumcases} {\label{eq:two_species_flocking_transition}}
		\sigma_{1/2}(k=0) = 0 \label{eq:two_species_flocking_transition_density}\\
		\sigma_{3/4/5/6}(k=0) = \frac{1}{2}  \Bigg[\big((g_{AA} + g_{BB})\,\frac{\rho_0}{2} - 2\big)  \pm \sqrt{- (\Omega_A + \Omega_B)^2}\nonumber \\
		\qquad \qquad \qquad \qquad \pm \sqrt{-(\Omega_A-\Omega_B)^2 + C(\{\Omega_a\}, \{g_{ab}\}) } \Bigg] \label{eq:two_species_flocking_transition_polarization}
	\end{subnumcases}
	with
	\begin{equation}
		\label{eq:two_species_flocking_function_c}
		\eqalign{
			\fl C(\{\Omega_a\}, \{g_{ab}\}) = \frac{\rho_0^2}{4} \, \Big[4\,g_{AB}\,g_{BA}+ (g_{AA} - g_{BB})^2 \Big]\\
			\quad + \frac{\sqrt{- (\Omega_A + \Omega_B)^2}}{\Omega_A + \Omega_B}  \, (g_{AA} - g_{BB}) \,\rho_0 \,  (\Omega_A - \Omega_B) ,
		}
	\end{equation}
	where we have chosen $\rho_0^A=\rho_0^B=\rho_0/2$ for simplicity.
	The first two growth rates \eref{eq:two_species_flocking_transition_density} vanish due the conservation of the particle density. The real parts of the other four eigenvalues can become positive for strong orientational alignment coupling, see \eref{eq:two_species_flocking_transition_polarization}. As in the one-species case, this generally indicates a flocking instability.
	The flocking instability becomes oscillatory for non-zero imaginary parts of the eigenvalues \eref{eq:two_species_flocking_transition_polarization}. 
	This generally happens when $\Omega_a \neq 0$. 
	
	In addition to (pure) flocking, we also observe pure MIPS and combined flocking and MIPS. To identify these instabilities, we employ the same characteristics as outlined for the one-species system in \sref{sec:one_species} and summarized in \tref{tab:one-species_eigenvalues_order_parameters}.
	
	We now discuss some specific features occurring in the two-species system. Maybe most intriguingly, the flocking instability can become oscillatory in the non-reciprocal binary system even in the absence of chirality, $\Omega_a=0$. Specifically, this happens for ``antagonistic'' inter-species couplings ($g_{AB}\,g_{BA}<0$) with 
	\begin{equation}
	\label{eq:two_species_chiral_phase}
		-4\,g_{AB}\,g_{BA}>(g_{AA}-g_{BB})^2 .
	\end{equation}
	In this case, the function $C$ defined in \Eref{eq:two_species_flocking_function_c} becomes negative, yielding a non-zero imaginary part in $\sigma_{3/4/5/6}(k=0)$. From a physical point of view, the antagonistic case describes a situation where particles of \mbox{species $A$} want to \mbox{(anti-)}align with particles of species $B$, but not vice versa. Hence, the species have opposite goals, such that they can never reach a configuration satisfying both. The resulting ``dynamical frustration'' renders the flocking instability oscillatory, and thus, time-dependent. We stress again that, in this time-dependent phase, termed ``chiral phase'' by Fruchart et al.~\cite{fruchart_2021_non-reciprocal_phase_transitions}, the continuous change of the flocking direction over time stems from non-reciprocal couplings and \textit{not} from the chirality of individual particles. 
	
	Additional insights regarding the flocking instability (\Eref{eq:two_species_flocking_transition_polarization}) can be obtained by looking at the respective eigenvectors (see \ref{sec:flocking_two_species}), which contain information about the orientation of the $A$ and $B$ flocks. The eigenvectors indicate whether the flocks are oriented parallel, anti-parallel or with a certain relative angle to each other. It turns out that for systems with equal intrinsic frequencies, $\Omega_A=\Omega_B$, a flocking instability with $g_{AB}\,g_{BA}>0$ always yields either exactly parallel or exactly anti-parallel flocks -- independent of whether the inter-species couplings are reciprocal or non-reciprocal. (This comprises the case of \textit{non}-chiral active particles considered by Fruchart et al.~\cite{fruchart_2021_non-reciprocal_phase_transitions}.) However, non-reciprocity becomes important for the relative orientation of the flocks as soon as the species do not share the same chirality. For instance, for opposite chiralities, $\Omega_A=-\Omega_B=\Omega$, the relative angle between the flocks indeed depends on the explicit values of the coupling strengths (see \ref{sec:flocking_two_species}).

	Despite these subtleties, we will continue using the term ``\mbox{(anti-)}flocking'' to describe the emergence of large-scale ordered motion in a broader sense -- including imperfect, i.e.~not exactly (anti-)parallel, orientations of flocks.

	\subsubsection{Stability diagram} 
	\label{sssec:two-species_stability_diagram}
\begin{figure}[h]
		\centering \includegraphics[width=0.6\textwidth]{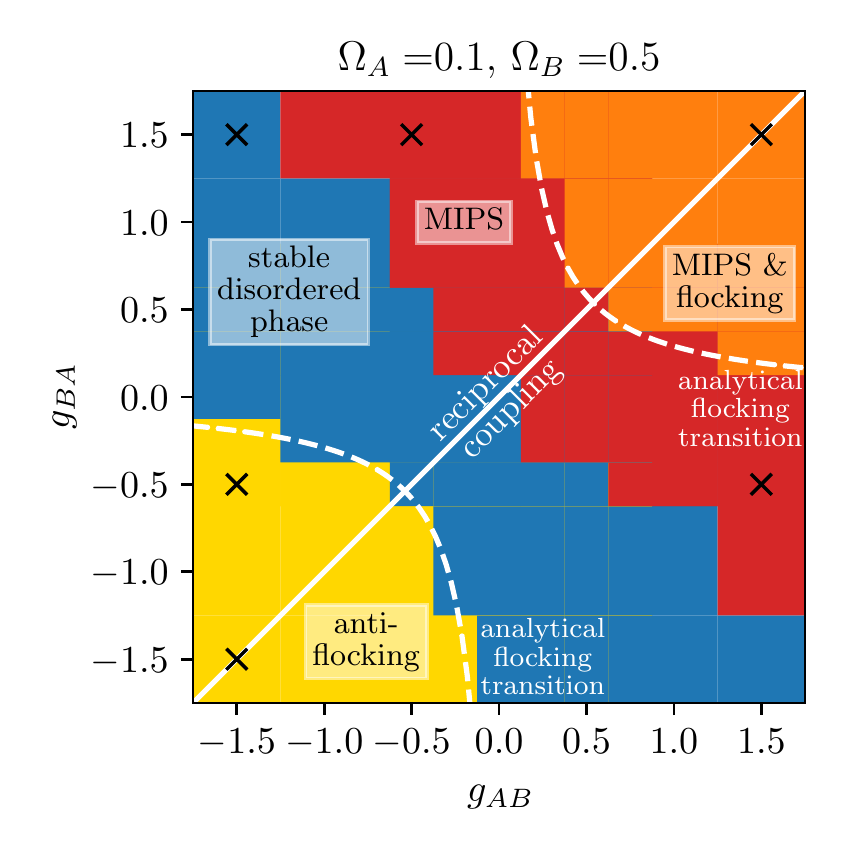}
		\caption{Linear stability diagram of the two-species chiral system obtained from \Eref{eq:two_species_linear_dynamics}. Depending on the inter-species coupling strengths $g_{AB}$ and $g_{BA}$, the system can exhibit four different collective states: stable disordered phase (blue), anti-flocking (yellow), MIPS (red) or MIPS combined with flocking (orange). Here, (anti-)flocking refers to flock \mbox{(anti-)}alignment, (whereby the flocks of species $A$ and $B$ are not necessarily perfectly (anti-)aligned). The relative angle between the flocks depends on the specific parameters. Reciprocal inter-species coupling ($g_{AB}=g_{BA}$) is indicated by the solid white line. Analytically determined (anti-)flocking regions are marked by dashed white lines. Black crosses indicate  parameters chosen in numerical continuum simulations in \sref{ssec:two_species_numerical_simulations}. Other parameters are $\Omega_A=0.1$, $\Omega_B=0.5$, $z=0.375\,{\rm Pe}_a/\rho^a_0$, and $g_{aa}=0.5$, ${\rm Pe}_a=1.5$, $\mathcal{R}=0.1$, $\rho_0^a=1$, and $D_{\rm t}=0.01$ with $a=A,B$.}
		\label{fig:phaseDiagramTwoSpecies}
	\end{figure}

	We now turn to the effect of non-reciprocal inter-species couplings on the linear stability of the disordered phase. To this end, we set the intrinsic frequencies exemplarily to the values $\Omega_A=0.1$ and $\Omega_B=0.5$. Further, we assume that particles within both species weakly align ($g_{AA}=g_{BB}=0.5$) and we chose a velocity-reduction parameter of $z=0.375\,{\rm Pe}_a/\rho^a_0$. In the one-species system with $\rho_0^A=\rho_0/2$ and $\Omega_A=0.1$, these parameters would result in a pure MIPS instability. Starting from this scenario, we now vary the inter-species coupling strengths $g_{AB}$ and $g_{BA}$. Our results regarding the linear stability of the system are summarized in the diagram in \fref{fig:phaseDiagramTwoSpecies}. The exemplarily shown real parts of the growth rates in \fref{fig:real_growth_rates_two_species} in \ref{sec:two-species_real_growth_rates} share much similarities with the one-species case (\fref{fig:growthRatesOneSpecies}).
	
	When $g_{AB}=g_{BA}$ (diagonal solid white line), the orientational coupling in the binary system is reciprocal. Therefore, when moving along the diagonal line, we only vary the strength of inter-species alignment ($g_{ab}>0$) or anti-alignment ($g_{ab}<0$). The system becomes non-reciprocal if $g_{AB} \neq g_{BA}$. Note that the stability diagram is indeed symmetric under the exchange $g_{AB} \leftrightarrow g_{BA}$. At $k=0$, this directly follows from the structure of Equations \eref{eq:two_species_flocking_transition} and \eref{eq:two_species_flocking_function_c}. For a more general discussion of this point, see \ref{sec:two_species_symmetry}.
	
	When particles of different species strongly anti-align ($g_{AB}<0$, $g_{BA}<0$), the linear stability analysis predicts a flocking instability (yellow, \fref{fig:real_growth_rates_two_species}(\textit{c}) and (\textit{d})). More specifically, following the discussion in \sref{ssec:two_species_linear_stability}, the corresponding eigenvectors predict anti-flocking with parameter-dependent relative angle between the $A$ and $B$ flocks. (Relative angles for exemplary parameter combinations are given in the subsequent \sref{ssec:two_species_numerical_simulations}.)
	Strong alignment ($g_{AB}>0$, $g_{BA}>0$), on the other hand, leads to flocking combined with MIPS (orange, \fref{fig:real_growth_rates_two_species}(\textit{a})), where the eigenvectors predict that flocks of both species move rather parallel. In both cases, it is possible to find the analytical flocking transition line from the growth rates \eref{eq:two_species_flocking_transition_polarization} at $k=0$ (dashed white line). In between the two \mbox{(anti-)}flocking regimes, the linear stability analysis predicts either a stable disordered phase (blue) or MIPS (red, \fref{fig:real_growth_rates_two_species}(\textit{b})). In particular, the disordered phase is stable for weak inter-species anti-alignment, while the MIPS emerges for weak alignment.
	
	To illustrate, in particular, the effect of non-reciprocity, let us consider the reciprocal situation generating MIPS combined with flocking (right upper corner in \fref{fig:phaseDiagramTwoSpecies}) as a starting point. Then, we gradually increase non-reciprocity by moving horizontally in the stability diagram towards smaller $g_{AB}$ while keeping the same $g_{BA}$. The linear stability analysis predicts that MIPS combined with a flocking instability first changes to pure MIPS until, eventually, the disordered phase is stabilized. This demonstrates the fact that non-reciprocity alone can dramatically change the character of the instability.
	
	The results discussed so far pertain to a specific choice of intrinsic frequencies. As we show in \ref{sec:two_species_effect_intrinsic_frequency}, a different choice slightly shifts the instability regions, while the qualitative picture remains.

	\begin{figure}[h]
		\centering \includegraphics[scale=0.8]{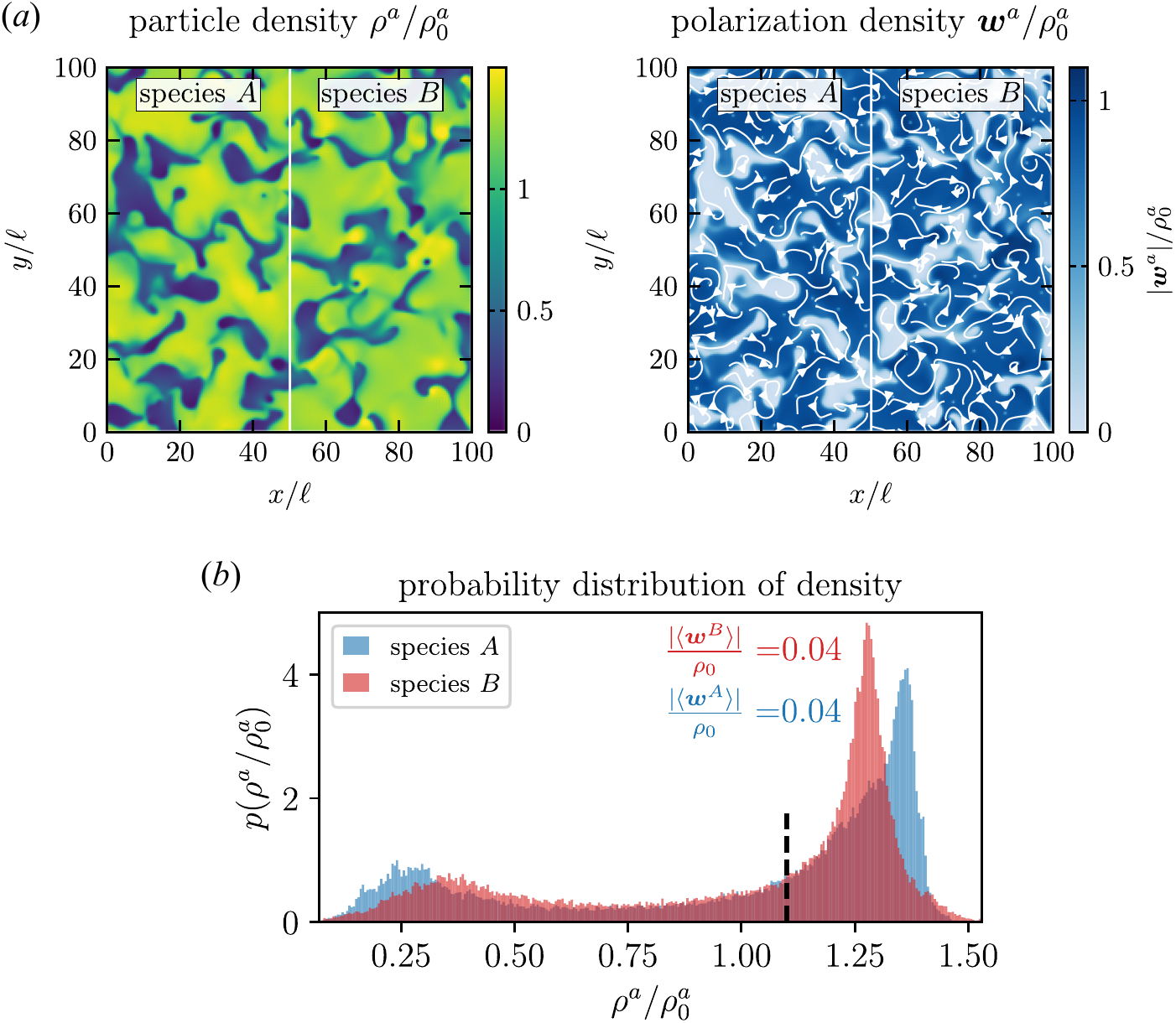}
		\caption{Numerical simulation results of MIPS combined with flocking in a reciprocal two-species system after the initial transient regime, $t>t_{\rm t}$. (\textit{a}) Representative snapshots of the time-dependent particle density $\rho^a(\bm{r},t)$ and polarization density field $\bm{w}^a(\bm{r},t)$. White arrows indicate the instantaneous, local direction of $\bm{w}^a$. (\textit{b}) Time-independent probability distributions $p(\rho^a)$ and absolute value of spatially-averaged polarizations $\vert \langle \bm{w}^a \rangle \vert$ in regions of large densities (larger than dashed vertical line). The parameters are $g_{AB}=g_{BA}=1.5$, $g_{aa}=0.5$, $z=0.375\,{\rm Pe}_a/\rho^a_0$, $\Omega_A=0.1$, $\Omega_B=0.5$, ${\rm Pe}_a=1.5$, $D_{\rm t}=0.05$, $\rho_0^a=1$, and $\mathcal{R}=0.1$ with $a=A,B$.}
		\label{fig:two_species_numerical_results_MIPS_flocking}
	\end{figure}

	\subsection{Numerical continuum simulations}
	\label{ssec:two_species_numerical_simulations}
	Performing numerical simulations of the full, non-linear hydrodynamic equations \mbox{\eref{eq:continuum_eq_density} -- \eref{eq:continuum_eq_polarization}} allows us to complement the linear stability analysis of the two-species system. Like in the one-species case, we choose parameter sets which lie well within the stability region of the respective phase transition in \fref{fig:phaseDiagramTwoSpecies} (black crosses). In particular, we choose a weak intra-species alignment of $g_{AA}=g_{BB}=0.5$. We now explore the effect of non-reciprocity by varying the inter-species coupling strengths $g_{AB}$, $g_{BA}$. Corresponding simulation videos can be found in the supplemental material.
	
	\begin{figure}[h]
		\centering \includegraphics[scale=0.8]{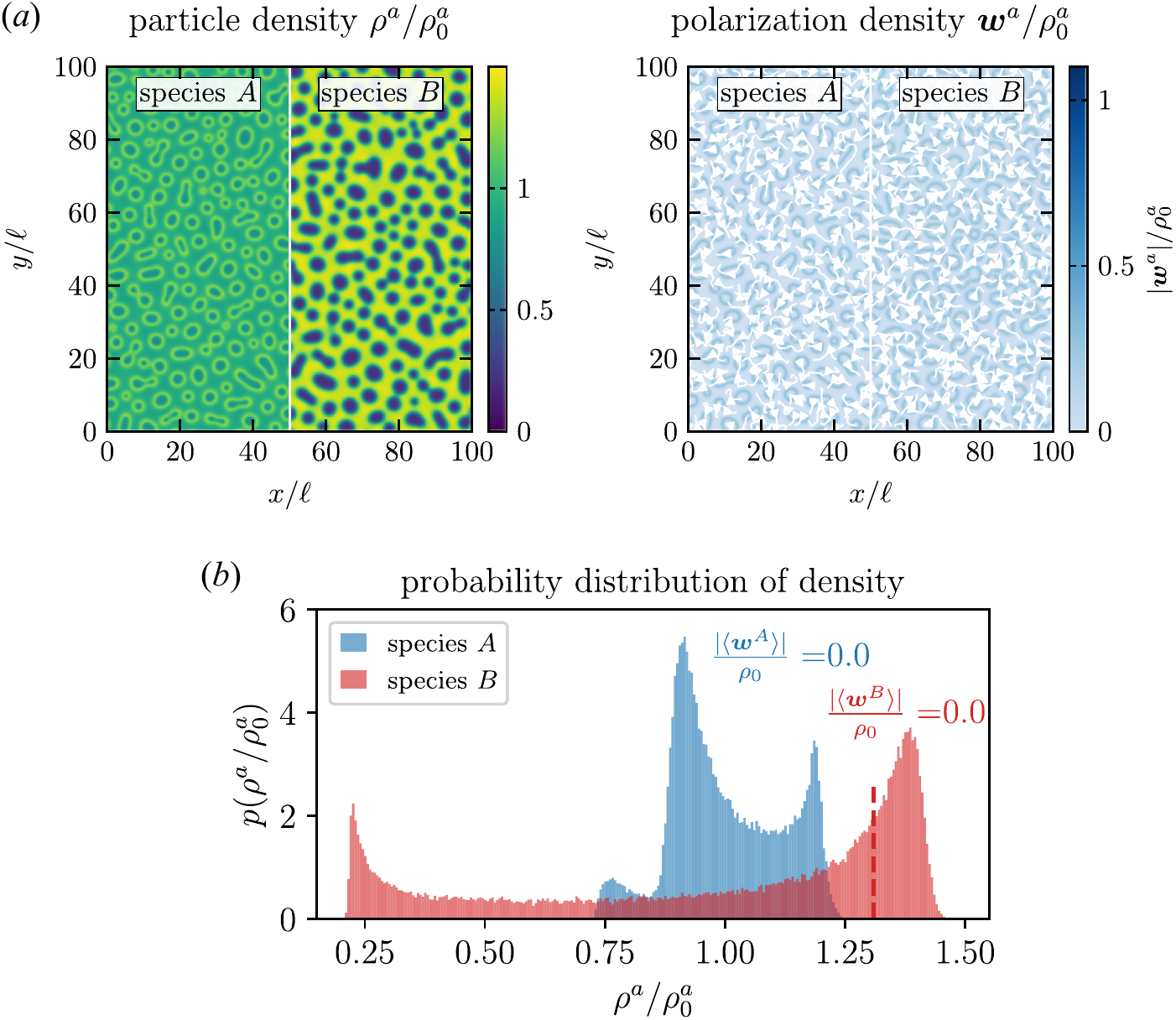}
		\caption{Numerical simulation results of MIPS in a non-reciprocal two-species system. (\textit{a}) Snapshots of the time-dependent particle density $\rho^a(\bm{r},t)$ and polarization density field $\bm{w}^a(\bm{r},t)$. The particle clusters grow in time. White arrows indicate the instantaneous, local direction of $\bm{w}^a$. (\textit{b}) Instantaneous probability distributions $p(\rho^a)$ and absolute value of spatially-averaged polarizations $\vert \langle \bm{w}^a \rangle \vert$ in regions of large densities (larger than dashed vertical line) for species $B$ and overall for species $A$. The parameters are $g_{AB}=-0.5$, $g_{BA}=1.5$, $g_{aa}=0.5$, $z=0.375\,{\rm Pe}_a/\rho^a_0$, $\Omega_A=0.1$, $\Omega_B=0.5$, ${\rm Pe}_a=1.5$, $D_{\rm t}=0.03$, $\rho_0^a=1$, and $\mathcal{R}=0.1$ with $a=A,B$.}
		\label{fig:two_species_numerical_results_MIPS}
	\end{figure}

	As a starting point, we consider a reciprocal mixture, which exhibits MIPS combined with a flocking instability (upper right cross in \fref{fig:phaseDiagramTwoSpecies}). Here, the two species only differ in their intrinsic frequencies and are coupled via strong reciprocal alignment of strength $g_{AB}=g_{BA}=1.5$. The representative snapshots of the time-dependent $\rho^a(\bm{r},t)$ and $\bm{w}^a(\bm{r},t)$ in \fref{fig:two_species_numerical_results_MIPS_flocking}(\textit{a}) show that both species form clusters of enhanced densities and non-vanishing polarization $\vert \bm{w}^a \vert$. Zoomed-in snapshots at two different times are shown in \fref{fig:two_species_zoomed-in_MIPS_flocking} in \ref{sec:two_species_zoomed-in_snapshots}. As in the one-species system, the clusters in the combined MIPS and flocking case constantly merge and break up. As a consequence, the average cluster size does not change much and the variances of the particle densities fluctuate around the constant values ${\rm var}(\rho^A)=0.17$ and ${\rm var}(\rho^B)=0.14$, whereby both species accumulate in the same regions. Although the intra-species alignment strengths are relatively weak ($g_{AA}=g_{BB}=0.5$), the strong inter-species alignment strengths ($g_{AB}=g_{BA}=1.5$) lead to an overall strong alignment of particles. Thus, the observed MIPS combined with flocking pattern is, as in the one-species system, reminiscent of ``interrupted motility-induced phase separation'' in aligning particle systems \cite{van_der_linden_2019_interrupted_mips}. Within the clusters, both species form rotating flocks. Each individual flock can be described by \Eref{eq:spatially_averaged_polarization}. The spatially-averaged rotation frequencies, $\Omega_{A,B\,{\rm flocks}} = 0.075<\Omega_A<\Omega_B$ (with variance ${\rm var}(\Omega)=0.015$) are the same for species $A$ and $B$ and stay constant after the initial transient regime. This indicates that the alignment couplings between the species prevent independent rotational motion, i.e.\ the flocks involve particles of both species. According to the stability analysis, the two polarization vectors enclose an angle of $\Theta_{\rm lin}=7.7$\textdegree \ (see \sref{ssec:two_species_linear_stability} and \ref{sec:flocking_two_species}), indicating that the flocks do not move exactly parallel to each other. The actually observed spatially-averaged angle is $\Theta_{\rm obs}(t)=4.5$\textdegree \ $\forall \ t$ after the transient regime ($t>t_{\rm t}$), which is remarkably close to the prediction. The time-independent probability distributions $p(\rho^a)$ of the particle densities $\rho^a$ in \fref{fig:two_species_numerical_results_MIPS_flocking}(\textit{b}) further reveal that the clusters of the more slowly rotating species ($A$) are larger than those of the faster rotating species ($B$). In fact, this complies with our previous observation that intrinsic rotation generally opposes MIPS (see \sref{ssec:one_species_effect_intrinsic_frequency}). 
	
	\begin{figure}[h]
		\centering \includegraphics[scale=0.8]{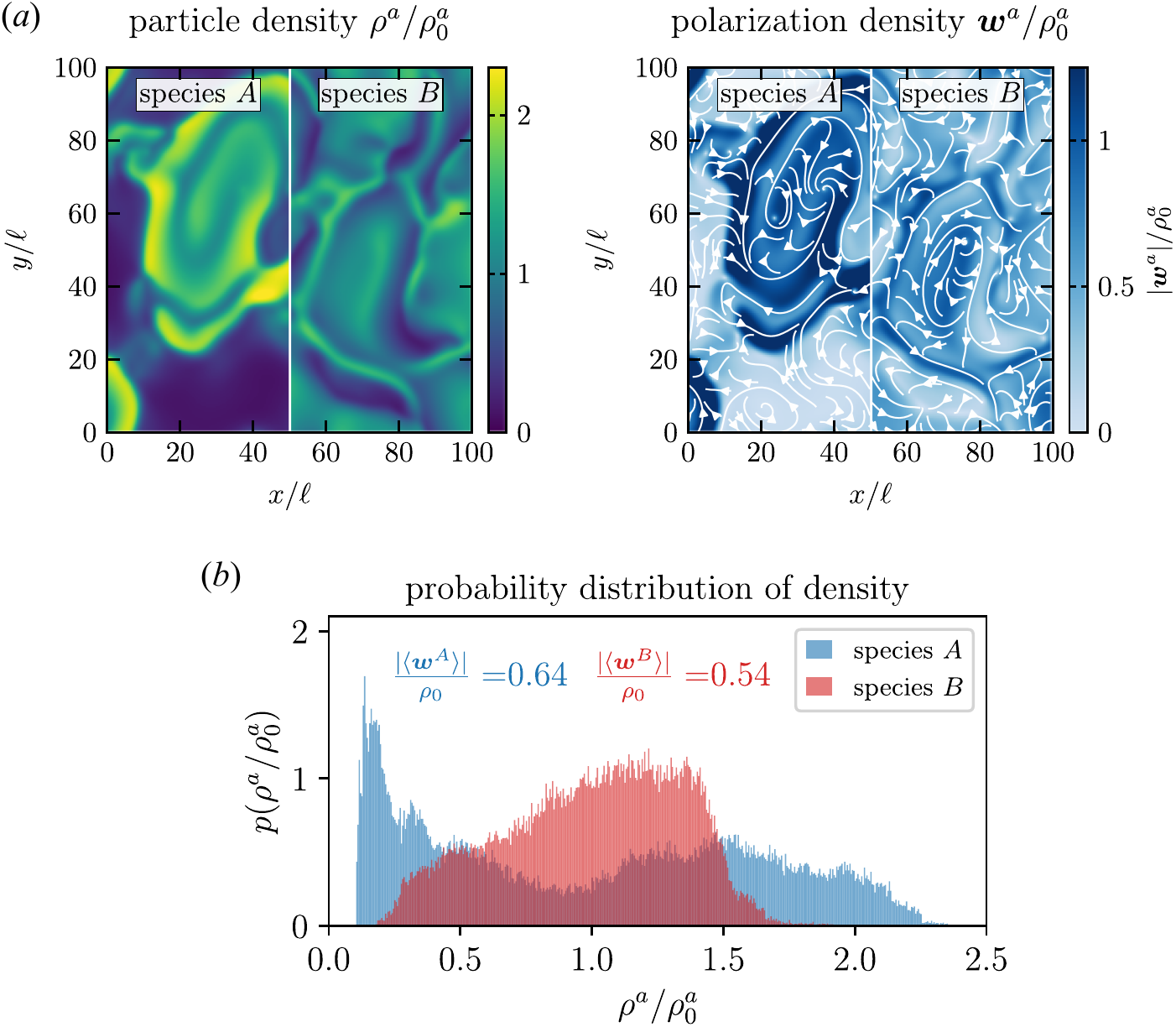}
		\caption{Numerical simulation results of anti-flocking in a reciprocal two-species system after the initial transient regime, $t>t_{\rm t}$. (\textit{a}) Snapshots of the time-dependent particle density $\rho^a(\bm{r},t)$ and polarization density field $\bm{w}^a(\bm{r},t)$. White arrows indicate the instantaneous, local direction of $\bm{w}^a$. \mbox{(\textit{b}) Time-independent probability} distributions $p(\rho^a)$ and absolute value of spatially-averaged polarizations $\vert \langle \bm{w}^a \rangle \vert$. The parameters are $g_{AB}=g_{BA}=-1.5$, $g_{aa}=0.5$, $z=0.375\,{\rm Pe}_a/\rho^a_0$, $\Omega_A=0.1$, $\Omega_B=0.5$, ${\rm Pe}_a=1.5$, $D_{\rm t}=0.3$, $\rho^a_0=1$, and $\mathcal{R}=0.1$ with $a=A,B$.}
		\label{fig:two_species_numerical_results_anti-flocking_reciprocal}
	\end{figure}

	Moving away from this reciprocal case by decreasing $g_{AB}$ relative to $g_{BA}$, the instability changes to pure MIPS (upper middle cross in \fref{fig:phaseDiagramTwoSpecies}). Corresponding snapshots presented in \fref{fig:two_species_numerical_results_MIPS}(\textit{a}) reveal that, indeed, both species form growing patterns with vanishing polarization. However, due to the non-reciprocal couplings between the species, the emerging patterns differ from each other. In particular, for the chosen parameters of antagonistic couplings with $g_{AB}=-0.5$ and $g_{BA}=1.5$, particles of species $A$ weakly anti-align with particles of species $B$, while the latter strongly align with particles of species $A$. The opposite goals combined with weak intra-species alignment prevents flocking of either species. Nevertheless, due to the stronger magnitude of $\vert g_{BA} \vert > \vert g_{AB} \vert$, one might assume that species $B$ ``wins'' the competition, resulting in enhanced alignment of species $B$ as compared to species $A$. As already weak particle alignment promotes MIPS (see \sref{ssec:one_species_effect_intrinsic_frequency}), this could be the reason for the enhanced cluster formation of species $B$, while particles of species $A$ accumulate at the edges of the clusters. This observation is also reflected in $p(\rho^a)$ in \fref{fig:two_species_numerical_results_MIPS}(\textit{b}). The peak at large densities of species $B$ indicates the cluster formation, whereas species $A$ is characterized by a more uniform density with a narrow peak at a slightly enhanced density. These numerical results match the linear stability analysis, which predicts the different cluster growth within the two species by means of the eigenvectors corresponding to the positive eigenvalues.

	A further decrease of $g_{AB}$ relative to $g_{BA}$ (upper left cross in \fref{fig:phaseDiagramTwoSpecies}) eventually stabilizes the disordered phase. Here, the antagonistic couplings, $g_{AB}=-g_{BA}=-1.5$, are of equally strong magnitude, preventing not only flocking but also MIPS of either species.
	
	\begin{figure}[h]
		\centering \includegraphics[scale=0.8]{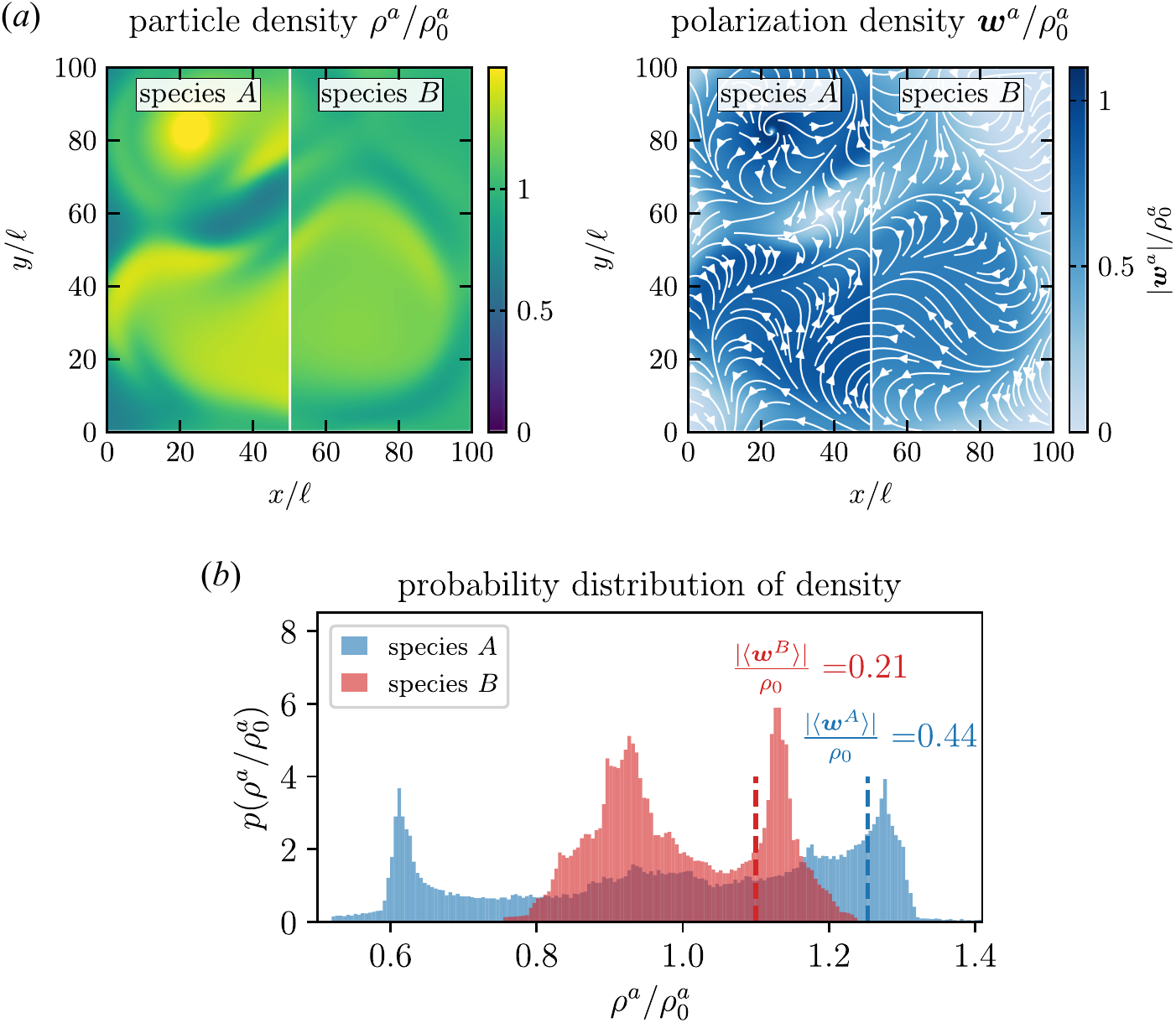}
		\caption{Numerical simulation results of anti-flocking in a non-reciprocal two-species system after the initial transient regime, $t>t_{\rm t}$. (\textit{a}) Snapshots of the time-dependent particle density $\rho^a(\bm{r},t)$ and polarization density field $\bm{w}^a(\bm{r},t)$. White arrows indicate the instantaneous, local direction of $\bm{w}^a$. \mbox{(\textit{b}) Time-independent probability} distributions $p(\rho^a)$ and absolute value of spatially-averaged polarizations $\vert \langle \bm{w}^a \rangle \vert$ in regions of large densities (larger than dashed vertical lines). The parameters are $g_{AB}=-1.5$, $g_{BA}=-0.5$, $g_{aa}=0.5$, $z=0.375\,{\rm Pe}_a/\rho^a_0$, $\Omega_A=0.1$, $\Omega_B=0.5$, ${\rm Pe}_a=1.5$, $D_{\rm t}=0.3$, $\rho_0^a=1$, and $\mathcal{R}=0.1$ with $a=A,B$.}
		\label{fig:two_species_numerical_results_flocking}
	\end{figure}

	Lastly, we turn to the anti-flocking regime (yellow region in \fref{fig:phaseDiagramTwoSpecies}). As a reference, we present in \fref{fig:two_species_numerical_results_anti-flocking_reciprocal}(\textit{a})representative snapshots of $\rho^a(\bm{r},t)$ and $\bm{w}^a(\bm{r},t)$ in the case of strong \textit{reciprocal} anti-alignment of strength $g_{AB}=g_{BA}=-1.5$ (lower cross in yellow region in \fref{fig:phaseDiagramTwoSpecies}). Zoomed-in snapshots at two different times are shown in \fref{fig:two_species_zoomed-in_reciprocal_flocking} in \ref{sec:two_species_zoomed-in_snapshots}. Both species form time-dependent patterns with non-vanishing polarization yet with different magnitude. Contrary to the previous cases of inter-species alignment ($g_{AB},g_{BA}>0$), where both species accumulated at the same places, particle clusters of different species seem to avoid each other in case of inter-species anti-alignment ($g_{AB},g_{BA}<0$). At the same time, particles of the slower rotating species $A$ accumulate into denser clusters than particles of species $B$. Even though the linear stability analysis does not predict MIPS, we observe cluster formation, which is, in accordance with our previous results (see \sref{ssec:one_species_effect_intrinsic_frequency}), weaker for larger intrinsic frequencies. The probability distributions $p(\rho^a)$ in \fref{fig:two_species_numerical_results_anti-flocking_reciprocal}(\textit{b}) further support these observations. Flocks of both species rotate with the same spatially-averaged frequency $\Omega_{A,B\,{\rm flocks}} = 0.19 > \Omega_A$ (variance ${\rm var}(\Omega)=0.02$), which is time-independent (after the initial transient regime). Thereby, the anti-alignment makes the two flocks move in rather anti-parallel direction. The linear stability analysis predicts a relative angle of $\Theta_{\rm lin}=172.3$\textdegree, which agrees very well with the observed angle of $\Theta_{\rm obs}(t)=172.8$\textdegree \ $\forall \ t>t_{\rm t}$.

	Introducing non-reciprocity while staying in the anti-flocking regime (upper cross in the yellow region in \fref{fig:phaseDiagramTwoSpecies}), we observe time-dependent patterns as shown in \fref{fig:two_species_numerical_results_flocking}(\textit{a}) and (\textit{b}) for $g_{AB}=-1.5$ and $g_{BA}=-0.5$. Zoomed-in snapshots at two different times are shown in \fref{fig:two_species_zoomed-in_non-reciprocal_flocking} in \ref{sec:two_species_zoomed-in_snapshots}. Clearly, the patterns differ from the reciprocal case considered in \fref{fig:two_species_numerical_results_anti-flocking_reciprocal} as both species now form larger clusters of enhanced densities. This is in line with the linear stability analysis, predicting only the growth of small-wavenumber perturbations as compared to the reciprocal case (see \fref{fig:real_growth_rates_two_species}(\textit{c}) and (\textit{d})). Still, particles of species $A$ strongly anti-align with particles of species $B$ ($g_{AB}=-1.5$), whereas now particles of species $B$ only weakly anti-align with particles of species $A$ ($g_{BA}=-0.5$). This results in significantly weaker flock formation for species $B$ than for species $A$ with $\vert \langle \bm{w}^B \rangle \vert < \vert \langle \bm{w}^A \rangle \vert$ (see \Eref{eq:magnitude_spatially_averaged_polarization}). The spatially-averaged and time-independent rotation frequencies $\Omega_{A,B\,{\rm flocks}} = 0.31 > 3\,\Omega_A$ (variance of $O(10^{-4})$) of species $A$ and $B$ flocks are considerably larger than the rotation frequency $\Omega_A$ of individual particles of species $A$ and larger than in the reciprocal anti-flocking case considered above. Also the relative angle of the formed flocks is affected: The observed angle of $\Theta_{\rm obs}(t)=166.9$\textdegree \ $\forall \ t>t_{\rm t}$ (predicted $\Theta_{\rm lin}=166.7$\textdegree) is less than in the reciprocal case. 
	
	Taken together, the results from the numerical continuum simulations underline the marked impact of non-reciprocity on the collective behavior of chiral active systems. Non-reciprocity can, first, change the general type of collective non-linear behavior and, second, also alter the emerging patterns within certain instability regimes.
	
	We close with two comments: The first one concerns the symmetry of the stability diagram in \fref{fig:phaseDiagramTwoSpecies}. Having seen the numerical simulation results, it is not surprising that although exchanging $g_{AB} \leftrightarrow g_{BA}$ does not affect the \textit{linear} stability of the system, the full, non-linear dynamics is not symmetric under the exchange (see also \ref{sec:two_species_symmetry}).
	
	The second point is related to the time-dependent phase termed ``chiral phase'' by Fruchart et al.~\cite{fruchart_2021_non-reciprocal_phase_transitions}. This phase (not to be confused with the rotating flocking phases in our system) emerges for strong enough antagonistic couplings (see \Eref{eq:two_species_chiral_phase}). To illustrate that our hydrodynamic model also captures these cases, numerical simulation results for a non-chiral system ($\Omega_A=\Omega_B=0$) with stronger intra-species alignment are shown in \ref{sec:chiral_phase_non-chiral_systems}.

	\section{Conclusion}
	\label{sec:conclusion}
	In this paper, we have studied the collective behavior of chiral active particles, interacting via volume exclusion and orientational couplings. In particular, we have allowed for non-reciprocal (anti-)alignment between active particles of different species.
	
	We have started from a particle-level description of the swimmers in terms of Langevin equations, in which an effective, density-dependent propulsion velocity accounts for the trapping of particles in crowded situations due to steric repulsion.
	We have then derived the corresponding coarse-grained description under the mean-field assumption and a scaling ansatz for higher-order orientational moments. The resulting hydrodynamic equations consist of a continuity equation for the (conserved) particle density field and a second time evolution equation for the (non-conserved) polarization density. These equations have been analyzed by linear stability analyses around the homogeneous, isotropic state, and by numerical solutions of the full, non-linear equations.
	
	We have first focused on the effect and interplay of the intrinsic frequency, steric repulsion, and reciprocal orientational couplings by studying the one-species system. Here, we found three different types of non-trivial collective behavior: flocking, MIPS, and flocking combined with MIPS. 
	Different to flocking, MIPS is strongly affected by the intrinsic frequency of the constituents. In particular, the intrinsic frequency generally opposes MIPS. However, this effect can to some extend be compensated by an increase of alignment strength, which promotes the emergence of MIPS. With this, our work presents the first hydrodynamic results for MIPS in systems of aligning chiral active particles. Despite the mean-field character of our hydrodynamic approach, the predictions qualitatively agree with particle-based simulations performed in earlier studies \cite{Liao_2018_circle_swimmers_monolayer}. 
	
	To explore the effect of non-reciprocal couplings between particles, we have then turned to a two-species system. 
	Our results demonstrate that non-reciprocity has indeed a significant impact on the collective dynamics. Consistent with recent field-theoretical results for non-reciprocal systems (e.g., \cite{You_Baskaran_Marchetti_2020_pnas,fruchart_2021_non-reciprocal_phase_transitions}), otherwise stationary instabilities can become oscillatory when couplings between particles are antagonistic. The (repulsive) system considered here thus adds an important example of an active soft-matter system with non-trivial time-dependent states, here generated by the interplay of non-reciprocity and chirality. Additionally, non-reciprocal inter-species couplings affect the relative orientation of the formed flocks. However, the most severe effect of non-reciprocity is that it can even change the general type of instability.
	
	In the present paper, we have mainly focused on a qualitative description of the emerging behavior in repulsive chiral active systems with non-reciprocal orientational couplings. As stated earlier in the paper (see \sref{ssec:one_species_numerical_simulations}), it would be desirable to complement our hydrodynamic results with particle-based simulations of the underlying Langevin equations. In this way one could avoid the (mean-field like) approximations in the derivation of the continuum equations, which would allow to study the characteristics of the individual phases and transitions, such as the active self-assembly into flocks, in more detail. Of course, the price to pay are much larger computational costs needed to find relevant parameter sets and analyzing the results. The present calculations could serve as a guideline for parameter sets and phenomena to be investigated. A further interesting question concerns the nature of the non-equilibrium transitions (or bifurcations in the continuum picture) separating different states. Work in these directions has just started (on the basis of simpler, non-reciprocal models \cite{frohoff-hulsmann_2021_Cahn-Hilliard_oscillatory,frohoff-hulsmann_2021_localized_states}).
	
	We also note that our findings are relevant for real chiral active mixtures such as anisotropic colloid systems \cite{campbell_2017_helical_paths_anisotropic_colloid}, bacteria close to walls \cite{di_leonardo_2011_swimming_bacteria_interface,lauga_2006_circle_swimming_bacteria,berg_1990_bacteria_swimming_circle,maeda_1976_circle_swimming_bacteria} or sperm cells \cite{friedrich_2007_chemotaxis_sperm_cells,riedel_2005_vortex_sperm_cell}. Due to naturally occurring heterogeneity and, in particular, couplings mediated through non-equilibrium environments, non-reciprocity is indeed pervasive in active matter systems. Moreover, as pointed out in \cite{Bowick_2022_symmetry_thermodynamcs_topology_active_matter}, non-reciprocity has far-reaching biological implications as it might be crucial in order to understand directed information transmission in living systems.

	\ack
	This work was funded by the Deutsche Forschungsgemeinschaft (DFG, German Research Foundation) -- Projektnummer 163436311 -- SFB 910.

	\appendix
	
	\section{Derivation of hydrodynamic equations from Langevin equations}
	\label{sec:derivation_hydrodynamic_equations_appendix}
	For the derivation of the hydrodynamic description from our Langevin equations in section \ref{ssec:coarse-grained_description}, we closely follow derivations presented in \cite{fruchart_2021_non-reciprocal_phase_transitions,Liebchen_2016_pattern_formation_chemically_interacting_active_rotors,liebchen_2017_collective_behavior_chiral_active_matter_pattern_formation_flocking, Levis_2019_activity_induced_synchronization}.
	
	As a first step, we derive from the Langevin equations \eref{eq:Langevin_r} and \eref{eq:Langevin_theta} a time-evolution equation for the stochastic ``fine-grained'' density referring to individual particles \cite{Dean_1996-Langevin_equation_interacting_particles},
	\begin{equation}
		g_{\alpha}(\bm{r},\theta,t) = \delta(\bm{r}-\bm{r}_{\alpha}(t)) \, \delta(\theta - \theta_{\alpha}(t)).
	\end{equation}
	Using It\^o's lemma and following \cite{Dean_1996-Langevin_equation_interacting_particles,fruchart_2021_non-reciprocal_phase_transitions}, we obtain its stochastic time evolution
	\begin{equation}
		\label{eq:stochastic_eq_fine_grained_PDF}
		\eqalign{
			\fl \frac{\partial}{\partial t}  g_{\alpha}(\bm{r},\theta,t) = - \nabla \cdot \Big\{g_{\alpha}(\bm{r},\theta,t)\, \Big(v_a\,\bm{p}(\theta) + \mu_r \int \int \sum_{\beta \neq \alpha} \bm{F}_{\rm{sr}}(\bm{r},\bm{r}')\,g_{\beta}(\bm{r}',\theta',t)\,{\rm d}\bm{r}'\,{\rm d}\theta'  + \bm{\xi}_{\alpha}(t)\Big)\Big\} \\
			- \partial_{\theta} \,\Big\{g_{\alpha}(\bm{r},\theta,t) \, \Big( \omega_a + \mu_{\theta} \int \int \sum_{\beta \neq \alpha} \mathcal{T}_{\rm al}^{ab}(\bm{r},\bm{r}',\theta,\theta')\, g_{\beta}(\bm{r}',\theta',t)\,{\rm d}\bm{r}'\,{\rm d}\theta' + \eta_{\alpha}(t)\Big) \Big\} \\
			+ \xi \, \nabla^2 \, g_{\alpha}(\bm{r},\theta,t)+ \eta \, \partial_{\theta}^2 \, g_{\alpha}(\bm{r},\theta,t) ,
		}
	\end{equation}
	where $\nabla$ and $\partial_{\theta}$ denote derivatives in space ($\bm{r}$) and orientation angle ($\theta$), respectively. We then sum over all ($N_a$) particles $\alpha$, take the ensemble average, and employ the mean-field approximation
	\begin{equation}
		\label{eq:mean-field_assumption}
		\fl f^{ab}(\bm{r},\bm{r}',\theta,\theta',t)  = \frac{1}{N_a\,N_b} \sum_{\alpha, \beta} \langle g_{\alpha}(\bm{r},\theta,t) \, g_{\beta}(\bm{r}',\theta',t) \rangle \approx f^a(\bm{r},\theta,t)\,f^b(\bm{r}',\theta',t)
	\end{equation}
	to approximate the two-particle PDF $f^{ab}(\bm{r},\bm{r}',\theta,\theta',t)$. 
	To treat the spatial integrals in \Eref{eq:stochastic_eq_fine_grained_PDF}, we assume that the coupling ranges $R_{r}$, $R_{\theta}$ are small such that only near-by particles interact. Changing then to the particle distance $d=\vert \bm{r}'-\bm{r} \vert$ as new integration variable, we can perform Taylor expansions of the terms inside the integrals up to first order around $d=0$. These expansions allow us to perform the spatial integration analytically. As a result we obtain the Fokker-Planck equation
	\begin{equation}
		\label{eq:Fokker-Planck_equation}
		\eqalign{
			\fl \frac{\partial}{\partial t} f^{a}(\bm{r},\theta,t) = - \nabla \cdot \Big\{f^{a}(\bm{r},\theta,t)\, \Big(v_a\,\bm{p}(\theta) + \frac{2\,\pi}{3}\, k \, \mu_r\, R_{r}^4 \, \sum_b N_b \, \nabla \,f^{b}(\bm{r},t) \Big)\Big\} \\
			- \partial_{\theta} \,\Big\{f^{a}(\bm{r},\theta,t) \, \Big( \omega_a + \pi \, \mu_{\theta} \,R_{\theta}^2 \int \sum_b N_b\, K_{ab} \, {\rm sin}(\theta'-\theta)\, f^b(\bm{r},\theta',t)\,{\rm d}\theta' \Big) \Big\} \\
			+ \xi \, \nabla^2 \, f^{a}(\bm{r},\theta,t)+ \eta \, \partial_{\theta}^2 \, f^{a}(\bm{r},\theta,t) .
		}
	\end{equation}
	Note that the same Fokker-Planck equation can be obtained by determining drift and diffusion coefficients using the Kramers-Moyal expansion of the distribution function, as described for general settings in \cite{risken_fokker-planck_1996}. Importantly, one would also need the same mean-field assumption \eref{eq:mean-field_assumption} as employed in this work.
	
	Following \cite{Bertin_2009_hydrodynamic_equations_self-propelled_particles}, the calculation of the remaining orientational integral in the Fokker-Planck equation~\eref{eq:Fokker-Planck_equation} and further derivation of the hydrodynamic equations greatly simplifies when we express the one-particle PDF in terms of its Fourier expansion with respect to the angle $\theta$, i.e.
	\begin{equation}
		f^a(\bm{r},\theta,t) = \frac{1}{2\,\pi} \, \sum_{n=-\infty}^{n=\infty} \hat{f}_n^a(\bm{r},t) \, {\rm e}^{-in\theta}.
	\end{equation}
	Relating the two-dimensional derivative $\nabla$ appearing in the Fokker-Planck equation~\eref{eq:Fokker-Planck_equation} to the complex quantity $\partial_z = \partial_x + i\,\partial_y$ with complex conjugate $\partial_{\overline{z}} = \partial_x - i \, \partial_y$ \cite{fruchart_2021_non-reciprocal_phase_transitions}, a lengthy but straightforward calculation yields the time evolution of the Fourier modes $\hat{f}^a_n(\bm{r},t) = \int_{-\pi}^{\pi} f^a(\bm{r},\theta,t) \, {\rm e}^{in\theta}\,{\rm d}{\theta} $, reading
	\begin{equation}
		\label{eq:Fourier_modes_time_evolution}
		\eqalign{
			\fl \partial_t \hat{f}_n^a = -\frac{1}{2} \left[ \partial_z \, (v_a\,\hat{f}_{n-1}^a) +  \partial_{\overline{z}}\,(v_a\, \hat{f}_{n+1}^a) \right] 	+ \frac{\pi}{3}\,k\, \mu_r \,R_{r}^4 \, \sum_b N_b \left[\partial_z \left\{ \hat{f}_n^a \, \partial_{\overline{z}} \, \hat{f}_0^b \right\} + \partial_{\overline{z}} \left\{  \hat{f}_n^a \, \partial_{z} \, \hat{f}_0^b \right\} \right] \\
			+\frac{R_{\theta}^2\,\mu_{\theta}\, \pi}{2}  \sum_b N_b \, K_{ab} \, n\,  \Big\{ \hat{f}_{n-1}^a\, \hat{f}_{1}^b - \hat{f}_{n+1}^a \, \hat{f}_{-1}^b \Big\}\\
			+ i\, \omega_a \, n\, \hat{f}_n^a - \xi \,  \partial_z\,\partial_{\overline{z}} \, \hat{f}_n^a - \eta \,  n^2 \, \hat{f}_n^a,
		}
	\end{equation}
	where we omitted the $(\bm{r},t)$-dependence of the Fourier modes $\hat{f}_n^a$. The identification of complex numbers with two-dimensional vectors allows us to relate the Fourier modes to moments of the one-particle PDF $f^a(\bm{r},\theta,t)$. In particular, we can identify the particle density (related to mode $n=0$)
	\begin{equation}
		\rho^a(\bm{r},t) = N_a \, \hat{f}^a_0(\bm{r},t),
	\end{equation}
	measuring the probability of finding a particle of species $a$ at position $\bm{r}$ and time $t$, and the polarization density (related to mode $n=1$)
	\begin{equation}
		\eqalign{
			\bm{w}^a(\bm{r},t) = N_a  \left(\matrix{{\rm Re}(\hat{f}_1^a) \cr {\rm Im}(\hat{f}^a_1)}\right) = N_a \int_{-\pi}^{\pi} f^a(\bm{r},\theta,t) \,\bm{p}(\theta) \, {\rm d}\theta,
		}
	\end{equation}
	describing the average orientation of particles of species $a$ via $\bm{w}^a/\rho^a$.
	
	The time evolution \eref{eq:Fourier_modes_time_evolution} of the Fourier modes $\hat{f}_n^a$ represents a hierarchy of equations, which requires a consistent closure scheme. Here we employ the scaling ansatz proposed in \cite{Bertin_2009_hydrodynamic_equations_self-propelled_particles, Bertin_2006_Boltzmann_hydrodynamic_self-propelled_particles}, which has already been used in numerous active matter systems, e.g.~\cite{fruchart_2021_non-reciprocal_phase_transitions,yllanes_2017_dissenters_to_disorder_a_flock,bricard_2013_emergence_macroscopic_directed_motion,peshkov_2014_boltzmann-ginzburg-landau}, including chiral active particle systems in \cite{Liebchen_2016_pattern_formation_chemically_interacting_active_rotors,liebchen_2017_collective_behavior_chiral_active_matter_pattern_formation_flocking, Levis_2019_activity_induced_synchronization}. The scaling ansatz assumes, first, that deviations from the isotropic state are so small that we can neglect moments of order $n=3$ or higher ($\hat{f}_n^a=0$ for $n\geq3$). Second, the scaling ansatz assumes that the nematic order parameter, which is related to the second mode $\hat{f}_2^a$, changes adiabatically, i.e.~$\partial_t \hat{f}_2^a=0$. 
	These assumptions allow us to express $\hat{f}_2^a$ solely in terms of the first Fourier mode $\hat{f}_1^a$. Specifically, we obtain from \Eref{eq:Fourier_modes_time_evolution},
	\begin{equation}
		\eqalign{\hat{f}_2^a = -\frac{i\,\omega_a + 2\,\eta}{8\,\eta^2 + 2\, \omega_a^2} \left(\frac{1}{2}\, \partial_z \,(v_a\,\hat{f}_1^a) - R_{\theta}^2\, \mu_{\theta} \, \pi \, \sum_b N_b \, K^{ab} \, \hat{f}_1^a \, \hat{f}_1^b \right).
		}
	\end{equation}
	As a result of the closure relation, the full dynamics of the one-particle PDF is reduced to the dynamics of the particle density and polarization density. The concrete coupled equations \eref{eq:continuum_eq_density}--\eref{eq:continuum_eq_polarization} for the particle density $\rho^a$ and polarization density $\bm{w}^a$ follow after some lengthy, but straightforward calculations and non-dimensionalization.

	\section{Full hydrodynamic equations in \sref{ssec:coarse-grained_description}}\label{sec:full_hydrodynamic_equations} 
	The full version of the time evolution of the polarization density (see \Eref{eq:continuum_eq_polarization}) is given by
	\begin{equation}
		\label{eq:continuum_eq_polarization_full}
		\eqalign{
			\fl \partial_t \bm{w}^a 
			= - \frac{1}{2} \, \nabla \,\big(v^{\rm eff}_a(\rho)\, \rho^a\big)  -\Omega_a \, \bm{w}^{a*} - \bm{w}^{a} + \sum_b g_{ab} \, \rho^a\, \bm{w}^b \\
			+  D_{\rm t}\,\nabla^2\,\bm{w}^a + \frac{v^{\rm eff}_a(\rho)}{4\,b_a} \, \nabla^2\,\Big(v^{\rm eff}_a(\rho)\,\Big\{2\,\bm{w}^a - \Omega_a\,\bm{w}^{a*} \Big\}\Big) \\
			- \sum_{b,c} \frac{2\, g_{ab}\,g_{ac}}{b_a} \, \Big[2 \, \bm{w}^a \, (\bm{w}^b \cdot \bm{w}^c) - \Omega^a \, \bm{w}^{a*} \, (\bm{w}^b \cdot \bm{w}^c) \Big] \\
			+ \sum_b \mathcal{R}_b \Big[ \bm{w}^a\, \nabla^2 \rho^b + (\nabla \rho^b) \cdot \nabla \bm{w}^a   \Big] \\
			- \frac{z}{4\,b_a} \, \big[\Omega_a \, \nabla^*\rho+ 2\,\nabla\rho\big] \cdot \big[\nabla \big(v^{\rm eff}_a(\rho)\,\bm{w}^a\big) - \nabla^* \big(v^{\rm eff}_a(\rho)\,\bm{w}^{a*} \big) \big]\\
			+ \sum_b \frac{g_{ab}}{2\,b_a} \Bigg\{ 2\,\Bigg[ \bm{w}^b \cdot \nabla \big(v^{\rm eff}_a(\rho)\, \bm{w}^a\big) + \bm{w}^{b*} \cdot \nabla \big(v^{\rm eff}_a(\rho)\, \bm{w}^{a*}\big)\\
			\qquad \qquad \qquad  -2\,v^{\rm eff}_a(\rho)\,\bm{w}^a \cdot \nabla \bm{w}^b - 2\,\bm{w}^b \, \nabla \cdot \big(v^{\rm eff}_a(\rho)\,\bm{w}^a\big) \\
			\qquad \qquad \qquad + 2\,v^{\rm eff}_a(\rho)\, \bm{w}^{a*} \cdot \nabla \bm{w}^{b*} + 2\,\bm{w}^{b*}\,  \nabla \cdot \big( v^{\rm eff}_a(\rho)\, \bm{w}^{a*}\big)  \Bigg]\\
			\qquad \qquad \ + \Omega_a \, \Bigg[ - \bm{w}^b \cdot \nabla \big(v^{\rm eff}_a(\rho)\, \bm{w}^{a*}\big) + \bm{w}^{b*} \cdot \nabla \big( v^{\rm eff}_a(\rho)\, \bm{w}^{a}\big) \\
			\qquad \qquad \qquad -2\,v^{\rm eff}_a(\rho)\, \bm{w}^a \cdot \nabla^* \bm{w}^b - 2\, \bm{w}^b\, \nabla^* \cdot \big(v^{\rm eff}_a(\rho)\, \bm{w}^a\big) \\
			\qquad \qquad \qquad + 2\, v^{\rm eff}_a(\rho)\, \bm{w}^{a} \cdot \nabla \bm{w}^{b*} + 2\, \bm{w}^{b*} \, \nabla \cdot \big( v^{\rm eff}_a(\rho)\,\bm{w}^a \big) \Bigg] \Bigg\} ,
		}
	\end{equation}
	where $v^{\rm eff}_a(\rho)={\rm Pe}_a- z\,\rho$ with $\rho=\sum_b\rho^b$, $b_a = 2(4+\Omega_a^2)$, $\bm{w}^*=(w_y, -w_x)^{\rm T}$, and $\nabla^*=(\partial_y, -\partial_x)^{\rm T}$.

	\section{Motility-induced phase separation in weakly-aligning non-chiral active systems}
	\label{sec:non-chiral_system_MIPS}
	A useful reference for the behavior of one-species chiral active systems discussed in \sref{sssec:one_species_results} is the case of zero intrinsic frequency ($\Omega_A=0$), which allows for analytically feasible expressions. The three eigenvalues of the non-chiral active system are 
	\begin{subnumcases} {\label{eq:one_species_non-chiral_growth_rate}}
			\sigma_1(k) = - 1 + g_{AA}\,\rho_0 -\left(\frac{v^2(\rho_0)}{16} + D_{\rm t}\right)\,k^2  \\
			\sigma_{2/3}(k) =  \frac{1}{2}\,\Big[- 1 + g_{AA}\,\rho_0 - \left( \frac{v^2(\rho_0)}{16} + \mathcal{R}\,\rho_0 + 2\,D_{\rm t}\right)\,k^2 \pm \sqrt{\mathcal{C}(k)} \Big] ,
	\end{subnumcases}
	where
	\begin{equation}
		\fl \eqalign{
			\mathcal{C}(k)= \left(1-g_{AA}\,\rho_0+\left(\frac{v^2(\rho_0)}{16}+\mathcal{R}\,\rho_0\right)\,k^2\right)^2\\
			\qquad -2\,k^2\,\left(v(\rho_0)\,({\rm Pe}-2\,z\,\rho_0)+2\,\mathcal{R}\,\rho_0\,(1-g_{AA}\,\rho_0)+\frac{v^2(\rho_0)}{8}\,\mathcal{R}\,\rho_0\,k^2\right)
		}
	\end{equation}
	and $v(\rho_0) = {\rm Pe} - z\,\rho_0$.

	\begin{figure}
		\centering
		\includegraphics[scale=0.8]{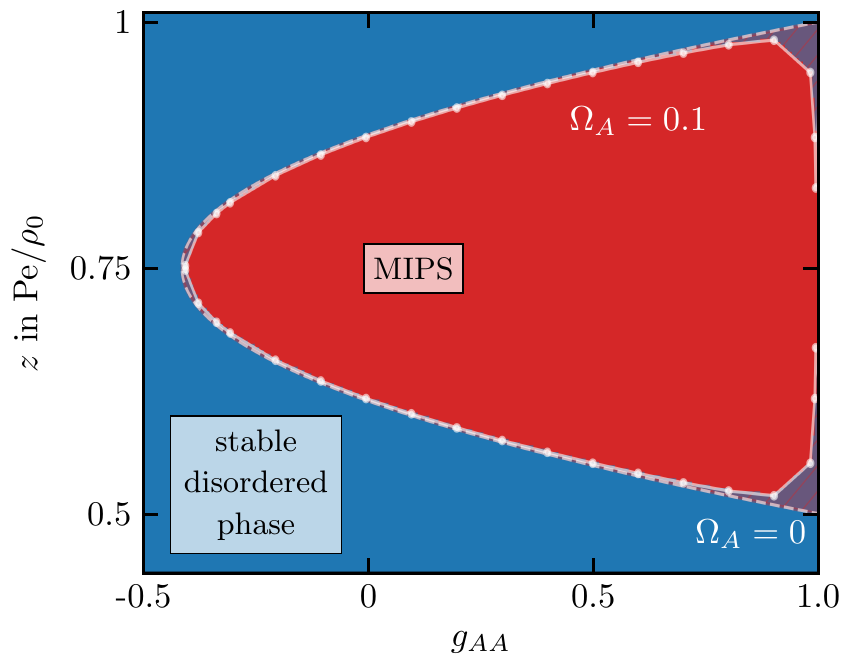}
		\caption{Occurrence of MIPS in systems of weakly aligning chiral active particles with intrinsic frequency $\Omega_A=0.1$ (white dots, analytically computed from \Eref{eq:one_species_linear_dynamics}) and non-chiral active particles, $\Omega_A=0$, from instability condition \eref{eq:non-chiral_swimmers_MIPS} (dashed line). Except for parameters very close to the flocking transition, differences are small. This changes as soon as particles rotate with an increased intrinsic frequency (see \sref{ssec:one_species_effect_intrinsic_frequency}). The results here pertain to $\rho_0=1$, $\rm{Pe}=1.5$, $\mathcal{R}=0.1$, and $D_{\rm t}=0$.}
		\label{fig:phaseDiagramOneSpeciesComparison}
	\end{figure}

	In the regime of weak alignment, i.e.~$g_{AA}\,\rho_0<1$, the eigenvalues $\sigma_{1}(k)$ and $\sigma_3(k)$ are negative at all $k$. In contrast, $\sigma_2(k)$ can be of typical MIPS form with $\sigma_2(k=0)=0$ and a positive maximum at a finite wave number. To see this, we expand eigenvalues the \eref{eq:one_species_non-chiral_growth_rate} up to second order in $k$, yielding
	\begin{subnumcases} {\label{eq:one_species_non-chiral_growth_rate_expansion}}
			\sigma_1(k) = g_{AA}\,\rho_0 - 1 -\Bigg(\frac{({\rm Pe-z\,\rho_0})^2}{16} + D_{\rm t}\Bigg)\,k^2  \\
			\sigma_{2}(k) = \Bigg(-(D_{t}+\mathcal{R}\,\rho_0) + \frac{({\rm Pe} - z\,\rho_0)\,({\rm Pe} - 2\,z\,\rho_0)}{2\,(g_{AA}\,\rho_0-1)}\Bigg)\,k^2 + O(k^3) \label{eq:one_species_non-chiral_growth_rate_expansion_sigma2}\\
			\sigma_{3}(k) =  g_{AA}\,\rho_0 - 1  - \Bigg(D_{\rm t} +({\rm Pe} - z\,\rho_0)\, \Bigg[ \frac{{\rm Pe} - 2\,z\,\rho_0}{2\,(g_{AA}\,\rho_0 -1)} + \frac{{\rm Pe}- z\,\rho_0}{16} \Bigg] \Bigg)\,k^2 + O(k^3).
	\end{subnumcases}
	From \eref{eq:one_species_non-chiral_growth_rate_expansion_sigma2} it follows that $\sigma_2(k)$ can become positive if
	\begin{equation}
		({\rm Pe}-z\,\rho_0)\,({\rm Pe}-2\,z\,\rho_0)+2\,(\mathcal{R}\,\rho_0+D_{\rm t})\,(1-g_{AA}\,\rho_0) < 0 .
	\end{equation}
	Thus, in the systems of non-chiral active particles with weak alignment interactions ($g_{AA}\,\rho_0<1$), we can expect pure MIPS to occur for velocity-reduction parameters $z_1>z>z_2$ with
	\begin{equation}
		\label{eq:non-chiral_swimmers_MIPS}
		z_{1/2} = \frac{3}{4} \, \frac{\rm Pe}{\rho_0} \pm \frac{1}{4\,\rho_0} \, \sqrt{{\rm Pe}^2 - 16\,(\mathcal{R}\,\rho_0+D_{\rm t})\,(1 - g_{AA}\,\rho_0)} .
	\end{equation}
	The expression on the right-hand side of \Eref{eq:non-chiral_swimmers_MIPS} resembles the one given by Ses\'e-Sansa et al.~\cite{elena_2021_phase_separation_self-propelled_disks} for active particles without alignment interactions (i.e.~$g_{AA}=0$). Here, we non-dimensionalize our equations with the rotational diffusion strength $\eta$, such that in our instability condition \eref{eq:non-chiral_swimmers_MIPS}, only the translational diffusion $\mathcal{R}\,\rho_0 + D_{\rm t}$ appears. Reference \cite{elena_2021_phase_separation_self-propelled_disks} uses a different re-scaling, such that $z_{1/2}$ depends on the product of rotational and effective translational diffusion coefficient ($D_r\,\mathcal{D}$).
	
	The stability diagram in \fref{fig:phaseDiagramOneSpeciesComparison} shows the comparison of the instability regions of non-chiral active particles, given by \Eref{eq:non-chiral_swimmers_MIPS}, and chiral active particles with $\Omega_A = 0.1$ (white dots). Note that for the computation of the instability region of chiral particles from eigenvalue equation \eref{eq:one_species_linear_dynamics}, we use the same procedure as for non-chiral particles (i.e.~expansion up to order $k^2$ to find values of $z$ for which the eigenvalue becomes positive). However, as analytical expressions of the eigenvalues are high-order polynomials, we only compute the instability conditions at exemplary values of $g_{AA}$. While the resulting differences in the stability regions are small for low intrinsic frequencies such as $\Omega_A=0.1$, larger values of $\Omega_A$ clearly affect MIPS. This is discussed in more detail in \sref{ssec:one_species_effect_intrinsic_frequency}.

	\section{(Anti-)Flocking in two-species system}
	\label{sec:flocking_two_species}
	Additional insights regarding the flocking instability (\Eref{eq:two_species_flocking_transition_polarization} in \sref{ssec:two_species_linear_stability}) in two-species systems can be obtained by looking at the respective eigenvectors. In particular, we here assume that the eigenvectors corresponding to the largest positive real part of the growth rates indicate the direction in the space of dynamical variables, in which perturbations grow the fastest. We further assume that the flock orientation of the individual species is given by $\hat{\bm{w}}^A_0 = \hat{\bm{w}}^A(k=0)$ and $\hat{\bm{w}}^B_0=\hat{\bm{w}}^B(k=0)$. Then, the relative angle $\Theta$ between $\hat{\bm{w}}^A_0$ and $\hat{\bm{w}}^B_0$ suggests whether the flocks are oriented parallel, anti-parallel or enclose a certain angle with each other.
	
	Some general remarks regarding the relative orientation of flocks can be deduced when we consider two exemplary choices of the intrinsic frequencies, namely $\Omega_A = \Omega_B$ and $\Omega_A = - \Omega_B$. For simplicity, we further set $g_{AA}=g_{BB}$ and $\rho_0^A=\rho_0^B=1$.
	
	When $\Omega_A=\Omega_B$, the eigenvalues at $k=0$ are given by (see \Eref{eq:two_species_flocking_transition_polarization})
	\begin{equation}
		\sigma_{3/4/5/6}(k=0) = \cases{
			-1 + g_{AA} + \sqrt{g_{AB}}\,\sqrt{g_{BA}} + i\, \Omega_A \\
			-1 + g_{AA} - \sqrt{g_{AB}}\,\sqrt{g_{BA}} + i\, \Omega_A \\
			-1 + g_{AA} + \sqrt{g_{AB}}\,\sqrt{g_{BA}} - i\, \Omega_A \\
			-1 + g_{AA} - \sqrt{g_{AB}}\,\sqrt{g_{BA}} - i\, \Omega_A .
		}
	\end{equation}
	Thus, a flocking instability occurs for inter-species couplings fulfilling the condition $\pm {\rm Re}\left(\sqrt{g_{AB}}\,\sqrt{g_{BA}}\right)> 1-g_{AA}$. The corresponding eigenvectors are
	\begin{equation}
		\fl \bm{v}_{3/4/5/6}(k=0) = (\hat{\rho}^A, \hat{w}_x^A, \hat{w}_y^A, \hat{\rho}^B, \hat{w}_x^B, \hat{w}_y^B)^{\rm T} = \cases{
			\left(0, i\, \frac{\sqrt{g_{AB}}}{\sqrt{g_{BA}}}, \frac{\sqrt{g_{AB}}}{\sqrt{g_{BA}}}, 0, i, 1  \right)^{\rm T} \\
			\left(0, - i\, \frac{\sqrt{g_{AB}}}{\sqrt{g_{BA}}}, - \frac{\sqrt{g_{AB}}}{\sqrt{g_{BA}}}, 0, i, 1  \right)^{\rm T} \\
			\left(0, - i\, \frac{\sqrt{g_{AB}}}{\sqrt{g_{BA}}}, \frac{\sqrt{g_{AB}}}{\sqrt{g_{BA}}}, 0, -i, 1  \right)^{\rm T} \\
			\left(0, i\, \frac{\sqrt{g_{AB}}}{\sqrt{g_{BA}}}, - \frac{\sqrt{g_{AB}}}{\sqrt{g_{BA}}}, 0, -i, 1  \right)^{\rm T}.
		}
	\end{equation}
	Since the total densities of both species are conserved quantities, the corresponding components are zero. To determine the relative angle between the flocks, we focus on the remaining components and calculate the angle between $({\rm Re}(\hat{w}_x^A), {\rm Re}(\hat{w}_y^A))^{\rm T}$ (flock $A$) and $({\rm Re}(\hat{w}_x^B), {\rm Re}(\hat{w}_y^B))^{\rm T}$ (flock $B$).
	The resulting relative angles of the growing eigenvectors are
	\begin{equation}
		\Theta_{\pm} = {\rm arccos}\left( \pm \frac{{\rm Re}\left(\frac{\sqrt{g_{AB}}}{\sqrt{g_{BA}}}\right)}{\sqrt{\vert \frac{g_{AB}}{g_{BA}}\vert}} \right),
	\end{equation}
	yielding
	\begin{equation}
		\cases{
			\Theta_+=0, \ \Theta_- = \pi & if $g_{AB}\,g_{BA} >0$ \\
			\Theta_{\pm} = \pi/2 & if $g_{AB}\,g_{BA} <0$.} 
	\end{equation}
	Hence, if $\Omega_A = \Omega_B$ and $g_{AB}\,g_{BA} >0$, the two flocks are either exactly parallel or exactly anti-parallel -- independent of whether the inter-species couplings are reciprocal or non-reciprocal. Only when the intra-species couplings are large, specifically, $g_{AA}=g_{BB}>1$, a flocking instability can occur for $g_{AB}\,g_{BA}<0$ with relative angle $\Theta=\pi/2$.
	These results are in accordance with findings by Fruchart et al.~\cite{fruchart_2021_non-reciprocal_phase_transitions}, who showed that for non-chiral active systems, inter-species couplings of the same sign always lead either to exactly parallel flocking or exactly anti-parallel anti-flocking.
	
	For opposite chiralities, $\Omega_A=-\Omega_B=\Omega$, the (degenerated) eigenvalues are given by (see \Eref{eq:two_species_flocking_transition_polarization})
	\begin{equation}
		\sigma_{3/4/5/6}(k=0) = -1 + g_{AA} \pm \sqrt{g_{AB}\,g_{BA} - \Omega_A} ,
	\end{equation}
	such that a flocking instability occurs as soon as \mbox{$\pm {\rm Re}\left(\sqrt{g_{AB}\,g_{BA} - \Omega_A^2}\right)> 1-g_{AA}$}. As in the previous case of same chirality, we focus on the real part of the eigenvector components to calculate the relative orientation of flock $A$ and flock $B$. For \mbox{$g_{AB}\,g_{BA}>0$}, the relative angle is given by
	\begin{equation}
	\label{eq:relative_angle_opposite_chirality}
		\fl \Theta_{\pm} = {\rm arccos}\left(\pm\,{\rm sign}(g_{AB},g_{BA})\,\frac{{\rm Re}\Big(\sqrt{g_{AB}\,g_{BA}-\Omega^2}\Big)}{\sqrt{\Omega^2+{\rm Re}\Big(\sqrt{g_{AB}\,g_{BA}-\Omega^2}\Big)^2}} \right) .
	\end{equation}
	\Eref{eq:relative_angle_opposite_chirality} reveals that in the case considered here (opposite chiralities), the angle between the flock directions can be different from $0$ or $\pi$ or $\pi/2$.
	The possibility of flocks, which move under a relative angle, has already been found in reciprocal chiral active mixtures \cite{levis_2019_chiral_active_mixtures, Levis_2019_activity_induced_synchronization}. 
	Here, we show that such a ``mutual flocking phase'' can also occur in non-reciprocal systems with opposite chiralities, as long as \mbox{$g_{AB}\,g_{BA}>0$}. However, for strongly non-reciprocal couplings with $g_{AB}\,g_{BA}<0$, the resulting relative angle of $\Theta=\pi/2$ is again not affected by the coupling strengths. 
	
	In fact, comparing the predicted relative angles between flocks with those obtained from numerical continuum simulations, we find good agreement at short times.
	
	\section{Real parts of growth rates in two-species system}
	\label{sec:two-species_real_growth_rates}

	\begin{figure}[H]
		\centering \includegraphics[width=0.8\textwidth]{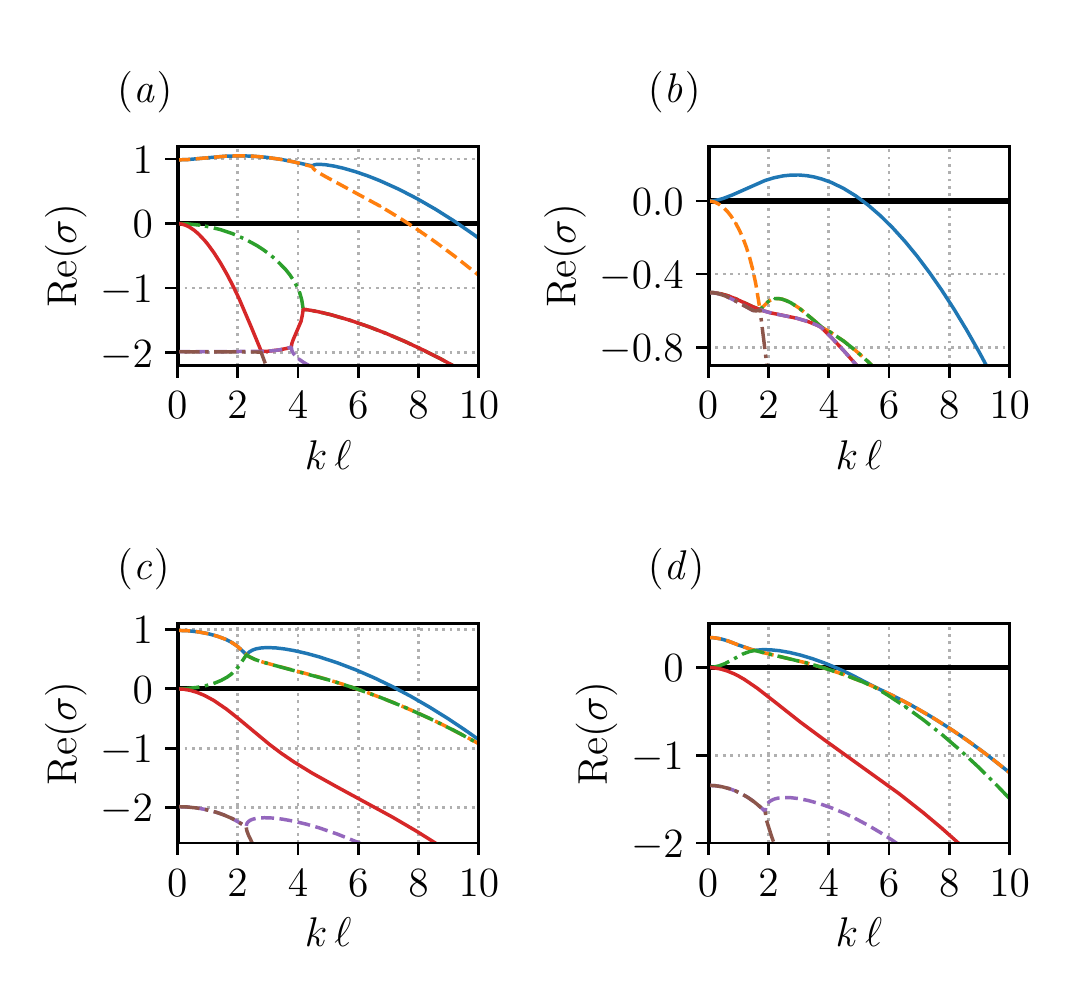}
		\caption{Real parts of the growth rates of emerging stability scenarios in the two-species chiral system with $\Omega_A= 0.1$, $\Omega_B=0.5$ for different inter-species coupling strengths $g_{AB}$, $g_{BA}$. Other parameters are set to $z=0.375\,{\rm Pe}_a/\rho^a_0$, $g_{aa}=0.5$, ${\rm Pe}_a=1.5$, $\mathcal{R}=0.1$, $\rho_0^a=1$, and $D_{\rm t}=0.01$ with $a=A,B$. (\textit{a}) MIPS combined with flocking for $g_{AB}=g_{BA}=1.5$. (\textit{b}) Pure MIPS for $g_{AB}=-0.5$, $g_{BA}=1.5$. (\textit{c}) Anti-flocking for $g_{AB}=g_{BA}=-1.5$. (\textit{d}) Anti-flocking for $g_{AB}=-1.5$, $g_{BA}=-0.5$.}
		\label{fig:real_growth_rates_two_species}
	\end{figure}

	In the two-species system different instability scenarios can be observed. The linear stability analysis (see \sref{ssec:two_species_linear_stability}) yields six growth rates, of which we here assume that the largest ones and corresponding eigenvectors determine the linear stability diagram in \fref{fig:phaseDiagramTwoSpecies}. Exemplary real parts of the growth rates for different scenarios (MIPS combined with flocking, pure MIPS, and anti-flocking) are shown in \fref{fig:real_growth_rates_two_species}.
	
	\section{Effect of intrinsic frequencies on linear stability of two-species system}
	\label{sec:two_species_effect_intrinsic_frequency}
	
	\begin{figure}[H]
		\centering \includegraphics[width=0.6\textwidth]{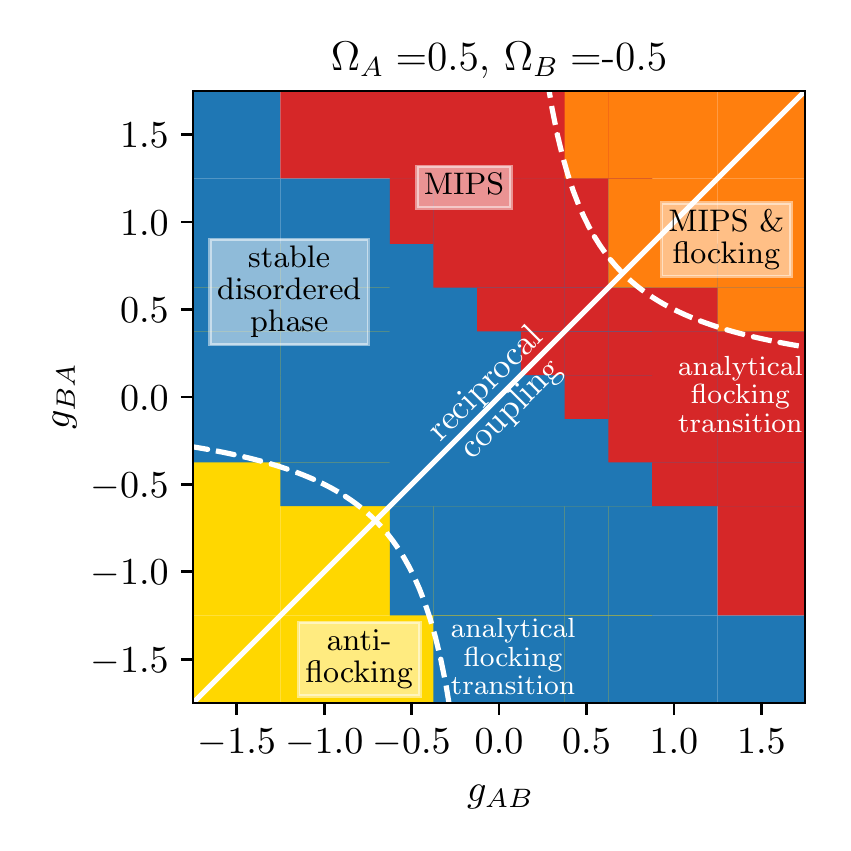}
		\caption{Linear stability diagram of the two-species chiral system with $\Omega_A= -\Omega_B=0.5$ obtained from \Eref{eq:two_species_linear_dynamics}. Depending on the inter-species coupling strengths $g_{AB}$ and $g_{BA}$, the system can exhibit four different collective states: stable disordered phase (blue), (imperfect) anti-flocking (yellow), MIPS (red) or MIPS combined with (imperfect) flocking (orange). Reciprocal inter-species coupling ($g_{AB}=g_{BA}$) is indicated by the solid white line. Analytically determined (anti-)flocking regions are marked by dashed white lines. Other parameters are set to $z=0.375\,{\rm Pe}_a/\rho^a_0$, $g_{aa}=0.5$, ${\rm Pe}_a=1.5$, $\mathcal{R}=0.1$, $\rho_0^a=1$, and $D_{\rm t}=0.01$ with $a=A,B$. 
		}
		\label{fig:phaseDiagramTwoSpecies_oppositeOmega}
	\end{figure}
	
	The intrinsic frequency of chiral active particles ($\Omega_A$, $\Omega_B$) has an impact on the stability diagram of the two-species system (in \sref{sssec:two-species_stability_diagram}). To illustrate this effect, we show in \fref{fig:phaseDiagramTwoSpecies_oppositeOmega} the stability diagram for mixtures with $\Omega_A = -\Omega_B = 0.5$. 
	It is seen that the system exhibits the same type of instabilities (MIPS, \mbox{(anti-)}flocking and MIPS combined with flocking) as in the case considered in \fref{fig:phaseDiagramTwoSpecies} ($\Omega_A=0.1$, $\Omega_B=0.5$). Closer inspection shows that the instability regions slightly differ from each other. Furthermore, the analytically determined flocking instability line, calculated from \Eref{eq:two_species_flocking_transition}, clearly depends on the intrinsic frequencies.

	\section{Zoomed-in snapshots of two-species systems}
	\label{sec:two_species_zoomed-in_snapshots}
	
\begin{figure}[h]
		\centering \includegraphics[width=\textwidth]{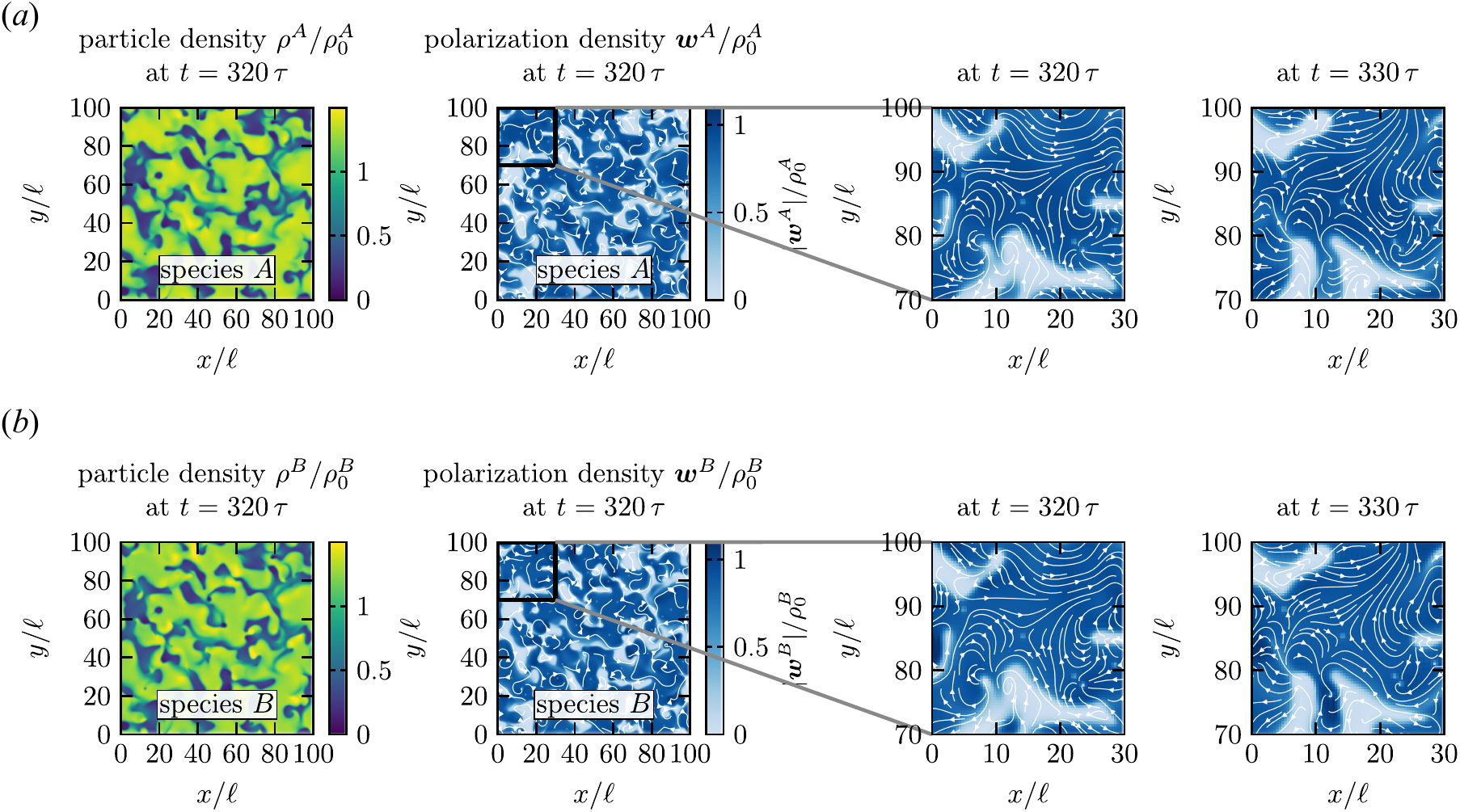}
		\caption{Numerical simulation results of MIPS combined with flocking in a reciprocal two-species system after the initial transient regime. Left: Representative snapshots of the time-dependent particle density $\rho^a(\bm{r},t)$ and polarization density field $\bm{w}^a(\bm{r},t)$. Right: Zoomed-in polarization density at two different times. White arrows indicate the instantaneous, local direction of $\bm{w}^a$. (\textit{a}) Species $A$. (\textit{b}) Species $B$. The parameters are $g_{AB}=g_{BA}=1.5$, $g_{aa}=0.5$, $z=0.375\,{\rm Pe}_a/\rho^a_0$, $\Omega_A=0.1$, $\Omega_B=0.5$, ${\rm Pe}_a=1.5$, $D_{\rm t}=0.03$, $\rho^a_0=1$, and $\mathcal{R}=0.1$ with $a=A,B$.}
		\label{fig:two_species_zoomed-in_MIPS_flocking}
	\end{figure}

\begin{figure}[h]
		\centering \includegraphics[width=\textwidth]{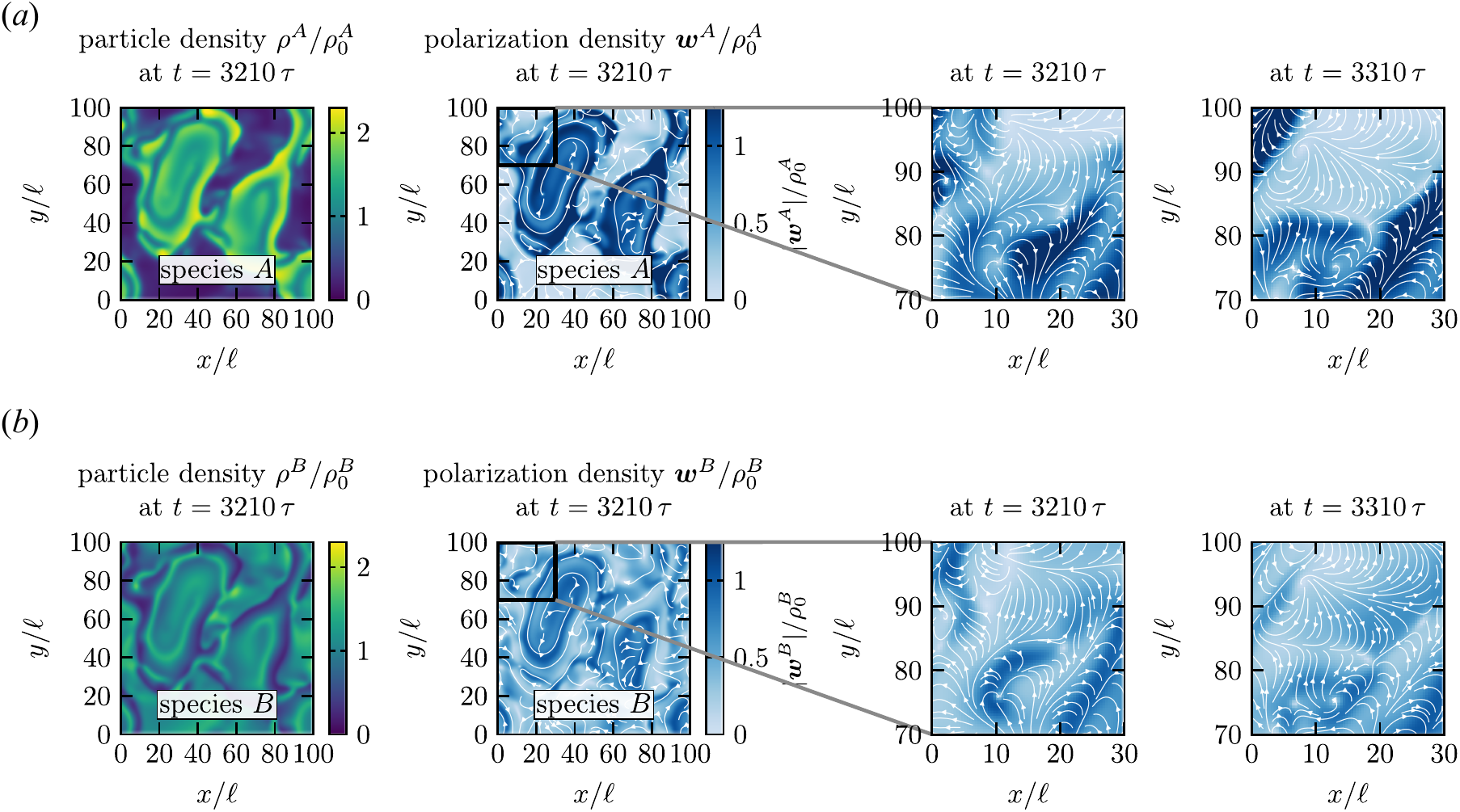}
		\caption{Numerical simulation results of anti-flocking in a reciprocal two-species system after the initial transient regime. Left: Representative snapshots of the time-dependent particle density $\rho^a(\bm{r},t)$ and polarization density field $\bm{w}^a(\bm{r},t)$. Right: Zoomed-in polarization density at two different times. White arrows indicate the instantaneous, local direction of $\bm{w}^a$. (\textit{a}) Species $A$. (\textit{b}) Species $B$. The parameters are $g_{AB}=g_{BA}=-1.5$, $g_{aa}=0.5$, $z=0.375\,{\rm Pe}_a/\rho^a_0$, $\Omega_A=0.1$, $\Omega_B=0.5$, ${\rm Pe}_a=1.5$, $D_{\rm t}=0.3$, $\rho^a_0=1$, and $\mathcal{R}=0.1$ with $a=A,B$.}
		\label{fig:two_species_zoomed-in_reciprocal_flocking}
	\end{figure}

\begin{figure}[H]
		\centering \includegraphics[width=\textwidth]{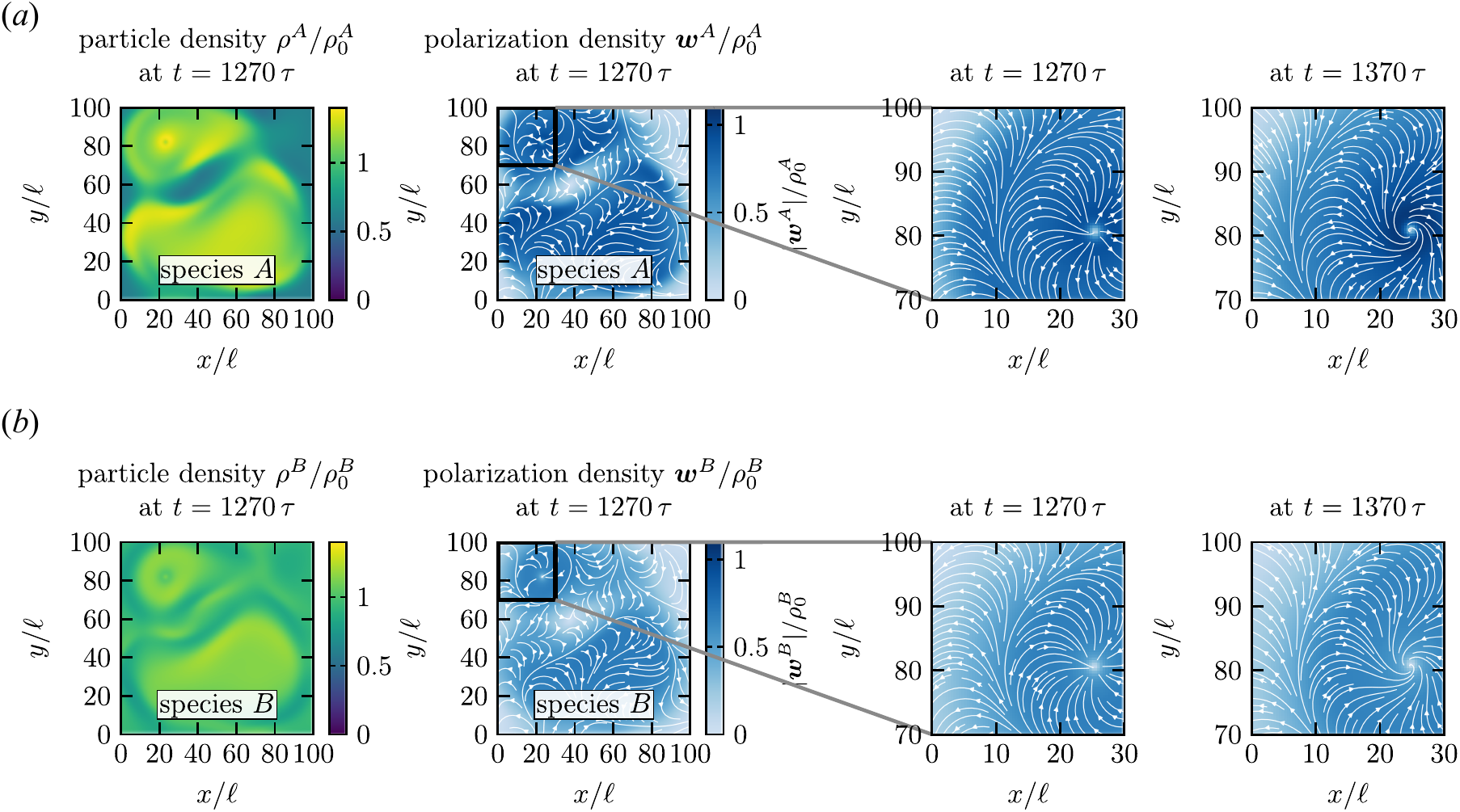}
		\caption{Numerical simulation results of anti-flocking in a non-reciprocal two-species system after the initial transient regime. Left: Representative snapshots of the time-dependent particle density $\rho^a(\bm{r},t)$ and polarization density field $\bm{w}^a(\bm{r},t)$. Right: Zoomed-in polarization density at two different times. White arrows indicate the instantaneous, local direction of $\bm{w}^a$. (\textit{a}) Species $A$. (\textit{b}) Species $B$. The parameters are $g_{AB}=-1.5$, $g_{BA}=-0.5$, $g_{aa}=0.5$, $z=0.375\,{\rm Pe}_a/\rho^a_0$, $\Omega_A=0.1$, $\Omega_B=0.5$, ${\rm Pe}_a=1.5$, $D_{\rm t}=0.03$, $\rho^a_0=1$, and $\mathcal{R}=0.1$ with $a=A,B$.}
		\label{fig:two_species_zoomed-in_non-reciprocal_flocking}
	\end{figure}

	In the two-species system, both species form different patterns with differently oriented flocks. To complement the descriptions of the various scenarios in \sref{ssec:two_species_numerical_simulations}, we here provide zoomed versions of non-vanishing polarization fields in the two-species system for the following three situations: reciprocal MIPS combined with flocking (\fref{fig:two_species_zoomed-in_MIPS_flocking}), reciprocal anti-flocking (\fref{fig:two_species_zoomed-in_reciprocal_flocking}), and non-reciprocal anti-flocking (\fref{fig:two_species_zoomed-in_non-reciprocal_flocking}). Since all emerging patterns are time-dependent due to the intrinsic frequency and non-reciprocal couplings between particles, we additionally show a zoomed version of the polarization field at a later time.

	\section{Symmetry of linear stability diagram in non-reciprocal two-species system}
	\label{sec:two_species_symmetry}
	
	\begin{figure}[h]
		\centering \includegraphics[scale=0.8]{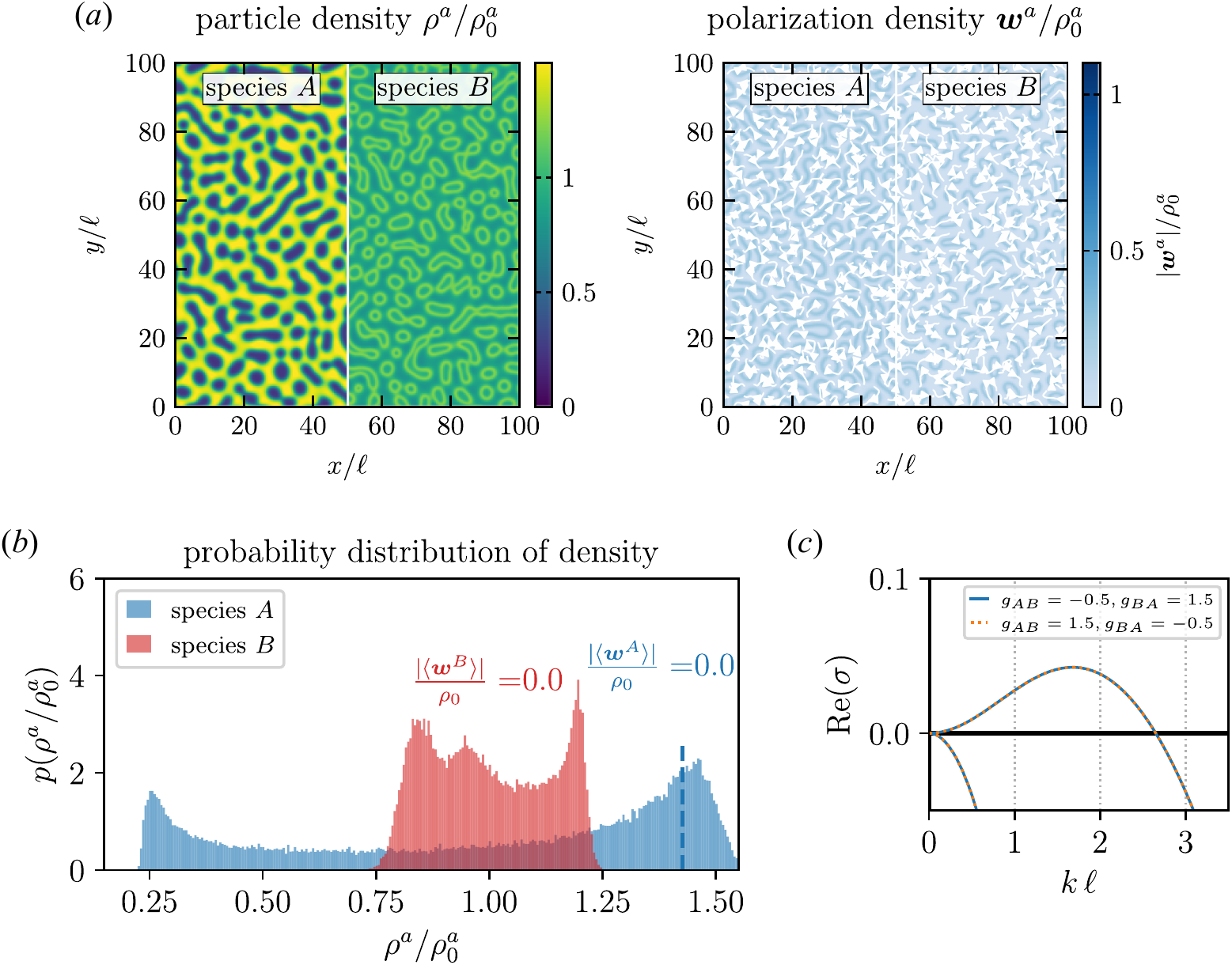}
		\caption{Numerical simulation results and growth rates in case of MIPS in a non-reciprocal two-species system. (\textit{a}) Snapshots of the time-dependent particle density $\rho^a(\bm{r},t)$ and polarization density field $\bm{w}^a(\bm{r},t)$. The particle clusters grow in time. White arrows indicate the instantaneous, local direction of $\bm{w}^a$. (\textit{b}) Probability distribution $p(\rho^a)$ and absolute value of spatially-averaged polarizations $\vert \langle \bm{w}^a \rangle \vert$ in regions of large densities (larger than right dashed vertical line) for species $A$ and overall for species $B$. Inter-species coupling strengths are $g_{AB}=1.5$, $g_{BA}=-0.5$. (\textit{c}) Largest real parts of growth rates for $g_{AB}\leftrightarrow g_{BA}$. Other parameters are $g_{aa}=0.5$, $z=0.375\,{\rm Pe}_a/\rho^a_0$, $\Omega_A=0.1$, $\Omega_B=0.5$, ${\rm Pe}_a=1.5$, $D_{\rm t}=0.03$, $\rho^a_0=1$, and $\mathcal{R}=0.1$ with $a=A,B$.}
		\label{fig:two_species_MIPS_symmetry}
	\end{figure}
	
	As mentioned in \sref{sssec:two-species_stability_diagram}, the instabilities predicted by our linear analysis are symmetric under the exchange $g_{AB}\leftrightarrow g_{BA}$, even though the chiral active particles rotate with different frequencies $\Omega_A\neq \Omega_B$. This is explicitly shown in \fref{fig:two_species_MIPS_symmetry}(\textit{c}), where we plot an example for the dominant growth rate. Clearly, ${\rm Re}(\sigma(k))$ is fully symmetric at all wave numbers.

	However, when performing simulations of the full, non-linear hydrodynamic equations \mbox{\eref{eq:continuum_eq_density} -- \eref{eq:continuum_eq_polarization}}, we see differences in the snapshots of both systems in \fref{fig:two_species_numerical_results_MIPS}(\textit{a}) and \fref{fig:two_species_MIPS_symmetry}(\textit{a}). While both systems form clusters with vanishing polarization, the probability distributions of particle densities differ (see \fref{fig:two_species_numerical_results_MIPS}(\textit{b}) and \fref{fig:two_species_MIPS_symmetry}(\textit{b})).

	\section{Numerical simulation of a time-dependent ``chiral'' phase in non-chiral, non-reciprocal two-species systems}
	\label{sec:chiral_phase_non-chiral_systems}	

\begin{figure}[H]
		\centering \includegraphics[scale=0.8]{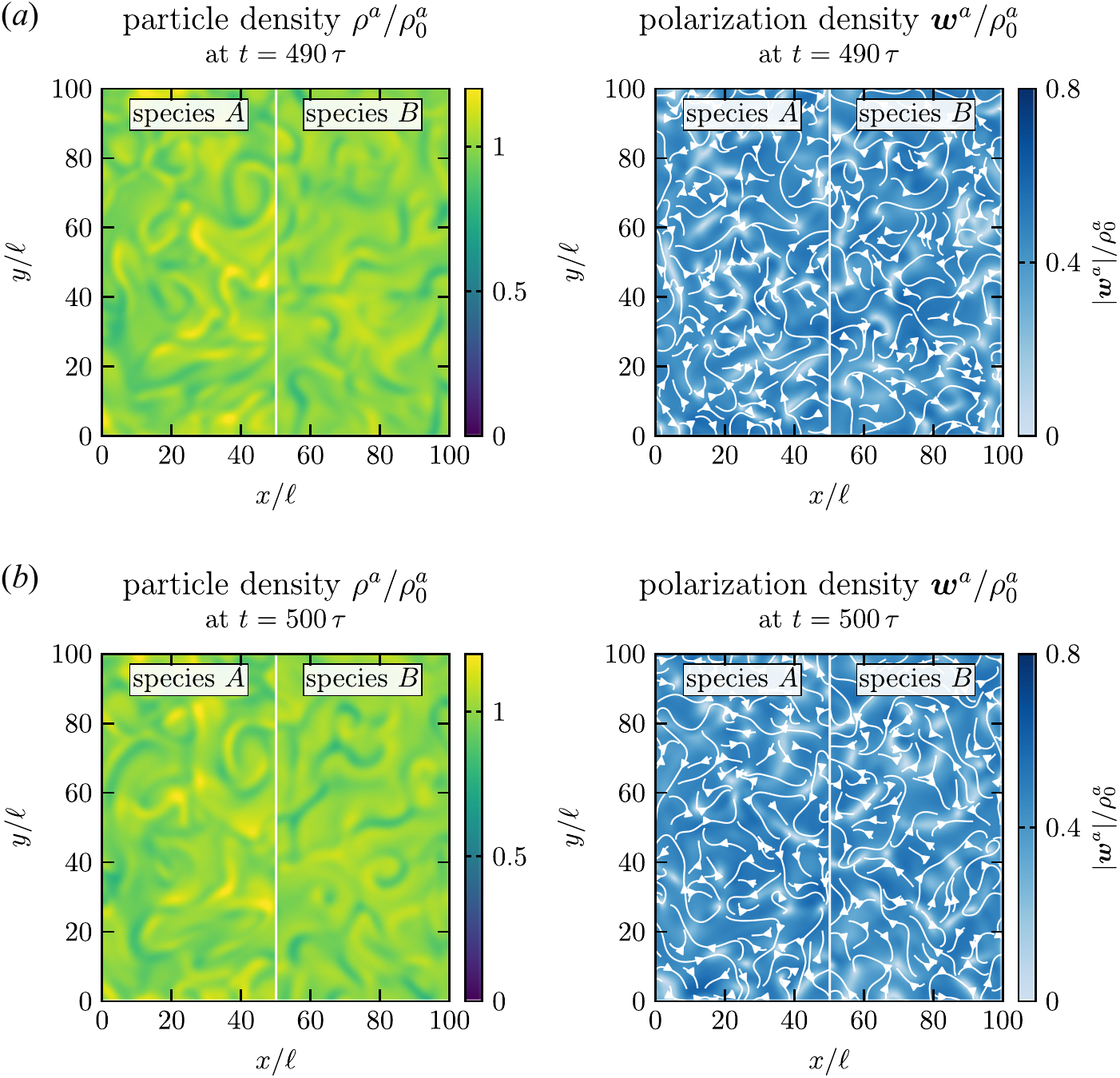}
		\caption{Numerical simulation results of MIPS combined with flocking in a non-chiral, non-reciprocal two-species system. Snapshots of the time-dependent particle density and polarization density field at time (\textit{a}) $t=490\,\tau$. (\textit{b}) $t=500\,\tau$. White arrows indicate the instantaneous, local direction of $\bm{w}^a$. The parameters are $g_{BA}=-g_{AB}=g_{aa}=1.5$, $z=0.375\,{\rm Pe}_a/\rho^a_0$, $\Omega_a=0$, ${\rm Pe}_a=1.5$, $D_{\rm t}=0.3$, and $\mathcal{R}=0.1$ with $a=A,B$.}
		\label{fig:two_species_numerical_results_non-chiral}
	\end{figure}

	In numerical simulations of our full hydrodynamic equations \eref{eq:continuum_eq_density} -- \eref{eq:continuum_eq_polarization}, we can observe a time-dependent ``chiral'' phase already in the absence of intrinsic rotation $\Omega_A=\Omega_B=0$ (i.e., in \textit{non}-chiral active systems), when couplings between species are antagonistic (see \Eref{eq:two_species_chiral_phase}). This intriguing phenomenon, predicted also in earlier field-theoretical studies \cite{fruchart_2021_non-reciprocal_phase_transitions}, is illustrated for the present system in \fref{fig:two_species_numerical_results_non-chiral}.
	As shown in the two snapshots in \fref{fig:two_species_numerical_results_non-chiral}(\textit{a}) and (\textit{b}), the resulting particle density and polarization density fields are time-dependent, whereby the direction of motion within the flocks continuously changes. In fact, after the initial transient regime, the flocks of both species move under a relative angle of $\Theta_{\rm obs}(t)=89.9$\textdegree \ $\forall \ t$, which is very well predicted by the linear stability analysis with $\Theta_{\rm lin}=90$\textdegree \ (see \ref{sec:flocking_two_species}).
	
	\section*{References}
	 \bibliographystyle{iopart-num-mod}
	 \providecommand{\newblock}{}

\end{document}